\documentstyle[12pt,epsfig,psfig]{article}

\newcommand{\newc}{\newcommand}
\newc{\gsim}{\lower.7ex\hbox{$\;\stackrel{\textstyle>}{\sim}\;$}}
\newc{\lsim}{\lower.7ex\hbox{$\;\stackrel{\textstyle<}{\sim}\;$}}
\newc{\gev}{\,{\rm GeV}}
\newc{\mev}{\,{\rm MeV}}
\newc{\ev}{\,{\rm eV}}
\newc{\kev}{\,{\rm keV}}
\newc{\tev}{\,{\rm TeV}}

\newc{\mz}{M_Z}
\newc{\mpl}{M_*}
\newc{\mw}{m_{\rm weak}}
\def\beq{\begin{equation}}
\def\eeq{\end{equation}}
\def\bea{\begin{eqnarray}}
\def\eea{\end{eqnarray}}
\newc{\ie}{{\it i.e.}}          \newc{\etal}{{\it et al.}}
\newc{\eg}{{\it e.g.}}          \newc{\etc}{{\it etc.}}
\newc{\cf}{{\it c.f.}}
\def\inv{^{\raise.15ex\hbox{${\scriptscriptstyle -}$}\kern-.05em 1}}
\def\lbar{{\lower.35ex\hbox{$\mathchar'26$}\mkern-10mu\lambda}} 

\def\om#1#2{\omega^{#1}{}_{#2}}

\let\<=\langle
\let\>=\rangle

\let\+=\uparrow

\let\om=\omega
\let\Om=\Omega

\begin{document}
\thispagestyle{empty}
\vspace*{.5cm}
\noindent
\hspace*{\fill}{\large DCPT-05/01}\\
\vspace*{0.5cm}

\begin{center}
{\Large\bf Bulk and Brane Decay of a (4+n)-Dimensional}\\[1mm]
{\Large \bf Schwarzschild-De-Sitter Black Hole:}\\[2mm]
{\Large \bf Scalar Radiation}

\bigskip \bigskip \medskip
{\large P. Kanti${}^1$, J. Grain${}^2$ and A. Barrau${}^2$
}\\[.5cm]
{\it ${}^{1}$Department of Mathematical Sciences, University of Durham,\\ South Road,
Durham DH1 3LE, UK}\\[.4cm]
{\it ${}^2$Laboratory for Subatomic Physics and Cosmology, Joseph Fourier University,
CNRS-IN2P3, 53, avenue de Martyrs, 38026 Grenoble cedex, France }
\\[1cm]

{\bf Abstract}\end{center}
\noindent
In this paper, we extend the idea that the spectrum of Hawking radiation can reveal
valuable information on a number of parameters that characterize a particular black
hole background -- such as the dimensionality of spacetime and the value of coupling
constants -- to gain information on another important aspect: the curvature of spacetime.
We investigate the emission of Hawking radiation from a $D$-dimensional
Schwarzschild-de-Sitter black hole emitted in the form of scalar fields, and
employ both analytical and numerical techniques to calculate greybody factors and
differential energy emission rates on the brane and in the bulk. The energy
emission rate of the black hole is significantly enhanced in the high-energy regime
with the number of spacelike dimensions. On the other hand, in the low-energy part of the
spectrum, it is the cosmological constant that leaves a clear footprint, through a
characteristic, constant emission rate of ultra-soft quanta determined by the
values of black hole and cosmological horizons. 
Our results are applicable to ``small" black holes arising in theories with an
arbitrary number and size of extra dimensions, as well as to pure 4-dimensional
primordial black holes, embedded in a de Sitter spacetime.

\newpage

\setcounter{page}{1}

\section{Introduction}

It has recently been pointed out that the hierarchy problem can be addressed in
an elegant geometrical way by assuming the existence of extra dimensions in the
Universe \cite{ADD, RS} (for some early works, see Ref. \cite{early}).
If only gravity is allowed to propagate in the $(4+n)$-dimensional
bulk while all other fields are confined to a (3+1)-dimensional brane, the
size of the extra dimensions can be as large as 1 mm and the related fundamental
Planck scale, $M_*$, can be as low as 1 TeV \cite{ADD}. A distinctive feature of the
scenario with Large Extra Dimensions \cite{ADD} is that, above the scale of
decompactification, gravitational interactions become strong, with the highly
suppressed 4-dimensional Newton's constant, $\kappa_4^2$, being replaced
by the significantly larger $D$-dimensional one, $\kappa_D^2$. This idea has
driven a considerable interest as it opens exciting possibilities for observing
strong gravitational phenomena possibly at the TeV scale.

A particularly exciting proposal is that of the creation of mini black holes
\cite{creation}, either at colliders \cite{colliders} or in high energy cosmic-ray
interactions \cite{cosmic} (for an extensive discussion of the phenomenological
implications as well as for a more complete list of references, see the recent
reviews \cite{Kanti, reviews}). If such black holes do indeed form during 
high-energy particle collisions, with center-of-mass energies greater than the
fundamental Planck scale $M_*$,
we expect them to evaporate through the emission of Hawking radiation, similarly to
their 4-dimensional counterparts - for a detailed discussion of the properties of
these microscopic black holes formed in a flat, higher-dimensional spacetime, see
Refs. \cite{Kanti, mp, admr}. These black holes remain, at least initially, attached
to our brane and emit Hawking radiation, in the form of elementary particles, both
in the bulk and on the brane. In order to avoid modification of the bulk gravitational
background, the self-energy of the brane is assumed to
be much smaller than the black hole mass. In addition, the mass of the produced black
hole must be, at least, a few times larger than the fundamental Planck scale $M_*$,
so that quantum gravity effects can be safely ignored, as for any classical object.

The emission of Hawking radiation from small black holes with horizon $r_H \ll L$,
where $L$ the size of the extra dimensions, formed in a flat,
higher-dimensional background, has been the subject of several works during the
last couple of years. In the case of a spherically-symmetric Schwarzschild-like
black hole, the task of determining the Hawking radiation emission rate has been
approached both analytically \cite{kmr1, Frolov1, kmr2} and numerically \cite{HK}.
In the first set of works, analytical formulae for the emission rate were derived,
but only under certain approximations that limited their validity either at the
low- or high-energy regime. The latter numerical analysis
\cite{HK}, though, demonstrated in the most exact way the dependence of the emission
rate on the fundamental parameters of the problem, i.e. the energy of the emitted
particle, its spin, and, last but not least, the dimensionality of spacetime. In
the case of a rotating, Kerr-like black hole in a flat higher-dimensional
spacetime, analytical formulae for the emission rate were again derived \cite{Frolov2,
IOP} but the results are only partial, being valid only at the low-energy regime,
for low angular momentum of the black hole, and for a specific dimensionality of
spacetime.

What these recent works -- focused on the emission from a higher-dimensional black hole
-- have revealed is the extremely important fact that the spectrum of Hawking radiation
encodes vital information about the structure of the geometrical background, a feature
that was not apparent in the related studies in the 4-dimensional spacetime
\cite{classics, page, sanchez}. Apart from the dimensionality of spacetime, the
radiation spectrum may give information on other aspects of the structure of the
particular black hole background emitting the radiation. For instance, in Ref.
\cite{Barrau}, it was demonstrated that, in the case of a higher-dimensional black
hole formed in the presence of the higher-derivative, stringy-inspired Gauss-Bonnet
term, the radiation spectrum depends also on the value of the coupling constant
of the Gauss-Bonnet term. 
 
During past years, significant observational evidence has been obtained favoring
a non-zero cosmological constant in the Universe \cite{cosmo}. The presence of a
non-trivial vacuum energy in the universe inevitably affects the formation of
black holes, and modifies accordingly the gravitational background around them. 
Since, as we mentioned above, the spectrum of Hawking radiation is highly sensitive
to the structure of the particular black hole background, the one emanating from
a black hole formed in a curved spacetime will also bear a strong dependence on 
the value of the cosmological constant. In 4 dimensions, the existing literature
consists of only a handful of works that have studied the emission of Hawking radiation
coming from a Schwarzschild-de-Sitter (SdS) black hole by either considering
two-dimensional toy models \cite{Mallett, Davies, Huang} or focusing on special
cases, like the degenerate case of a SdS spacetime where the black hole and
cosmological horizon coincide \cite{Bousso}. Surprisingly enough, in a higher
number of dimensions, no work up to now has investigated the exact form of the Hawking
radiation spectrum of a SdS black hole, with all the activity having been focused on
the study of the pair creation of black holes \cite{pair-creation}, thermodynamical
aspects of radiation via tunneling \cite{Medved}, or the perturbations of the
higher-dimensional spacetime \cite{ishibashi,Maeda} and the associated quasinormal
frequencies \cite{quasi} (for related studies in 4 dimensions, see \cite{quasi-4D}).

This article aims at filling the aforementioned gap in the literature by deriving the
exact form of the generalized Hawking radiation spectrum (including the greybody factors)
of a Schwarzschild black hole  embedded in a $D$-dimensional de Sitter space-time.
In this way, the exact dependence of the radiation spectrum on all the parameters
of the theory will be revealed, and the effects of both the dimensionality and the
bulk cosmological constant of the higher-dimensional spacetime on the emitted
radiation will be determined. From a theoretical point of view, the study of 
non-asymptotically flat spacetimes is well motivated by the de Sitter (dS)
and Anti-de Sitter (AdS)\,/\,conformal field theory (CFT) correspondence, and 
the need to generalize, for a curved background, techniques and principles,
developed for a flat spacetime, is obvious. From the observational point of view,
the prospect of obtaining information on the topological structure of our spacetime,
including its curvature, from the Hawking radiation emitted from small black holes --
an effect possibly observable in near or far-future experiments -- increases further
the importance of this work.

A discussion of the approximations made during our analysis should be added here.
In order to avoid any significant back reaction to the gravitational background 
due to the change in the black hole mass after the emission of a particle, we
must assume that $\omega \ll M_{BH}$. Since the emission spectrum is peaked around
the black hole temperature, it is sufficient to assume instead that $T_H \ll M_{BH}$.
As we will see in the next section, the temperature
of a Schwarszchild-de Sitter black hole is lower than the one of a Schwarzschild black
hole, therefore, if the latter satisfies the aforementioned constraint, the former will
also do. By using the relation between the horizon and mass for a $(4+n)$-dimensional
Schwarzschild black hole, this constraint translates to
$M_{BH} \gg M_*$, that is the black hole mass must be much bigger than the fundamental
Planck scale. The same bound guarantees that quantum effects are small since such a
black hole can be safely considered as a classical object. The frequency of the emitted particles
is also bounded from below: in order for these quanta to be considered as $(4+n)$-dimensional,
not only the black hole horizon, but also their wavelength must be smaller than the
size of the extra dimensions, or equivalently $\omega \gg 1/L$. Throughout our analysis,
we will be assuming that the mass of the black hole is indeed much higher than $M_*$;
however, the lower bound on the frequency will be temporarily ignored for the sake of
presenting complete emission spectra valid at all energy regimes - this bound will be
re-instated and its importance will be discussed in the final discussion of our results
at the end of our paper. 

The outline of our paper is as follows: in the next section, we describe the general
framework for our analysis and briefly discuss the properties of the higher-dimensional
SdS black hole. In section 3, we concentrate on the emission of Hawking radiation from a
decaying SdS black hole on the brane. We start by considering the line-element describing
the-induced-on-the-brane gravitational background, and then we derive the master equation
for propagation of fields with arbitrary spin in the 4-dimensional geometry. In this
work, we focus on the emission of Hawking radiation in the form of scalar fields, and
the corresponding cross-section, or``greybody factor", for emission on the brane is
determined, first analytically, in the high- and low-energy limits, and then numerically,
in terms of the total number of dimensions and the value of the bulk cosmological constant. 
The existence of extra dimensions in nature have a strong effect on the emission
cross-section, similar to the one found in the case of an asymptotically-flat
Schwarzschild black hole; in addition, the presence of the cosmological constant modifies
significantly the behaviour of the same quantity both at low and high energies. 
In section 4, we turn to the emission of Hawking radiation, in the form of scalar fields,
in the bulk: we provide again analytical results  for the ``bulk" greybody factor, in the
low- and high-energy regime, and then exact, numerical ones; both sets of results prove
the same strong dependence of the cross-section, for emission in the bulk, on the
cosmological constant and number of extra dimensions, as on the brane. In Section 5,
the final energy emission rates for the ``brane" and ``bulk" channels are calculated,
and the exact Hawking radiation spectra are presented. We comment on the role that
the two fundamental background parameters -- the cosmological constant and dimensionality
of spacetime -- play in the formation of the spectra at all energy regimes, and also
to the magnitude of the relative emission rate for ``bulk" and ``brane" emission.
The latter quantity is necessary to address the question of the amount of radiation
lost in the space transverse to the brane and thus the amount of energy available for
emission on the brane. We finish with a summary of our results and conclusions,
in section 6.

\section{General framework}

We start our analysis by presenting the line-element that describes a higher-dimensional,
neutral, spherically-symmetric black hole that arises in the presence of a positive
cosmological constant. The line-element of the so-called Schwarzschild-de-Sitter
(SdS) black hole was first derived in \cite{Tang}, and has the form
\beq
ds^2 = - h(r)\,dt^2 + \frac{dr^2}{h(r)} + r^2 d\Omega_{2+n}^2,
\label{bhmetric}
\eeq 
where
\beq
h(r) = 1-\frac{\mu}{r^{n+1}} - \frac{2 \kappa_D^2\,\Lambda\,r^2}{(n+3) (n+2)}\,.
\label{h-fun}
\eeq
The above line-element describes a black hole living in a $D$-dimensional spacetime, where
$n$ is the number of extra, spacelike dimensions
that exist in nature ($D=4+n$), and $\Lambda$ the positive cosmological constant in the
bulk. The parameter $\mu$ is related to the ADM mass of the black hole through the
relation \cite{mp} 
\beq
\mu=\frac{2 \kappa^2_D M_{BH}}{(n+2)\,A_{2+n}}\,, \qquad 
A_{2+n}=\frac{2 \pi^{(n+3)/2}}{\Gamma[(n+3)/2]}\,,
\eeq
where $A_{2+n}$ is the area of a unit $(2+n)$-dimensional sphere. In addition,
$d\Omega_{2+n}^2$ describes the corresponding line-element of the 
($2+n$)-dimensional unit sphere, and is given by
\begin{equation}
d\Omega_{2+n}^2=d\theta^2_{n+1} + \sin^2\theta_{n+1} \,\biggl(d\theta_n^2 +
\sin^2\theta_n\,\Bigl(\,... + \sin^2\theta_2\,(d\theta_1^2 + \sin^2 \theta_1
\,d\varphi^2)\,...\,\Bigr)\biggr)\,.
\label{unit}
\end{equation}
In the above, $0 <\varphi < 2 \pi$ and $0< \theta_i < \pi$, for 
$i=1, ..., n+1$ -- notice that due to the assumed spherical symmetry of the problem,
$n$ additional azimuthal coordinates $\theta_i$ have been introduced to describe
the $n$ compact extra dimensions. Finally, in Eq. (\ref{h-fun}), the parameter
$\kappa_D^2=1/M_*^{2+n}$ stands for the $(4+n)$-dimensional Newton's constant.

Due to the presence of the positive cosmological constant $\Lambda$ in the bulk, the
higher-dimensional spacetime is not asymptotically-flat. The spacetime has a true 
curvature singularity at $r=0$, and the equation
\beq
h(r_i) = 1-\frac{\mu}{r_i^{n+1}} - \frac{2 \kappa_D^2\,\Lambda\,r_i^2}
{(n+3) (n+2)}=0\,,
\label{horizon}
\eeq
will yield, in principle, $(n+3)$ roots $r_i$ which correspond to $(n+3)$ 
horizons for this spacetime. However, due
to the positivity of $r$, only two real, positive roots emerge, the largest one
($r_C$) corresponding to the Cosmological horizon, and the smallest one ($r_H$) to
the Black Hole event horizon. The metric function $h(r)$ is therefore interpolating
between two zeros, one at $r=r_H$ and one at $r=r_C$. The behaviour in between
is easily found by looking at the first derivative, $h'(r)$: this function becomes
zero only at the point
\beq
r_{0} = \biggl[\,\frac{(n+1) (n+2) (n+3) \mu}{4 \kappa^2_D \Lambda}\,\biggr]^{1/(n+3)}\,,
\label{r0}
\eeq
that is easily found to correspond to a global maximum. Therefore, $h(r)$ increases
for $r>r_H$, reaches a maximum at $r=r_0$, and decreases again towards zero for
$r_0<r<r_C$.

The temperature of the black hole is given in terms of the surface gravity at the
location of the horizon\,\footnote{In section 5, a more accurate definition of the 
temperature of a black hole embedded in a de Sitter spacetime will be given - for
the purpose of our analysis up to section 5, the above definition is sufficient.}, i.e.
\begin{equation}
T_H = \frac{k_H}{2 \pi} = \frac{1}{4\pi r_H}\,\Bigl[\,(n+1)- \frac{2\kappa^2_D \Lambda}
{(n+2)}\,r_H^2\,\Bigl]\,.
\label{temp-BH}
\end{equation}
In the above, we have used the relation that follows from Eq. (\ref{horizon}), i.e.
\beq
\mu = r_i^{n+1}\,\biggl[\,1- \frac{2\kappa^2_D \Lambda}
{(n+2)(n+3)}\,r_i^2\,\biggr]\,,
\label{mu}
\eeq
evaluated at $r_i=r_H$, to eliminate the parameter $\mu$ from the expression for the
temperature of the black hole. The temperature of the universe is given by a 
similar expression in terms of the surface gravity of the cosmological horizon
\begin{equation}
T_C = \frac{k_C}{2 \pi} = -\frac{1}{4\pi r_C}\,\Bigl[\,(n+1)-\frac{2\kappa^2_D \Lambda}
{(n+2)}\,r_C^2\,\Bigl]\,,
\end{equation}
where care has been taken to ensure that $T_C$ is positive. Since $r_C>r_H$, the 
temperature of the universe will always be smaller than that of the black hole,
$T_C<T_H$.

As any black hole with a non-vanishing temperature, the Schwarzschild-de-Sitter
black hole will emit Hawking radiation \cite{hawking} in the form of elementary particles. 
The emission will take place in the higher-dimensional spacetime (\ref{bhmetric})
with the corresponding flux spectrum, i.e. the number of particles emitted per unit
time, given by
\begin{equation}
\label{flux}
\frac{dN^{(s)}(\omega)}{dt} = \sum_{j} \sigma^{(s)}_{j,n}(\omega)\,
{1 \over \exp\left(\omega/T_{H}\right) \pm 1} 
\,\frac{d^{n+3}k}{(2\pi)^{n+3}}\,.
\end{equation}
The above formula is a direct generalization of the corresponding four-dimensional
expression \cite{hawking} for a higher number of dimensions. In the above, $s$ is
the spin of the emitted degree of freedom and $j$ its angular momentum quantum number.
The spin statistics factor in the denominator is $-1$ for bosons and $+1$ for fermions.
The corresponding power spectrum, i.e. the energy emitted per unit time by the black
hole, can be easily found by combining the number of particles emitted with the
amount of energy they carry. It is given by
\begin{equation}
\frac{dE^{(s)}(\omega)}{dt} = \sum_{j} \sigma^{(s)}_{j,n}(\omega)\,
{\omega  \over \exp\left(\omega/T_{H}\right) \pm 1}\,
\frac{d^{n+3}k}{(2\pi)^{n+3}}\,.
\label{power}
\end{equation}
For massless particles, $|k|=\omega$ and the phase-space integral reduces to an integral
over the energy of the emitted particle $\omega$. In what follows we are going to 
focus on the emission of massless particles, or on particles with a rest mass much
smaller than the temperature of the black hole. Note that, since $T_H>T_C$, there
will always be a net flow of energy from the black hole towards the universe.

Both expressions, Eqs. (\ref{flux}) and (\ref{power}), contain an additional factor,
$\sigma^{(s)}_{j,n} (\omega)$, which stands for the cross-section for the emission of
a particle from a black hole, or alternatively, the ``greybody factor". In a blackbody
radiation spectrum, this factor is constant, and equal to the area of the emitting body.
In the case of a black hole, however, it depends on the energy of the emitted particle,
its spin and its angular momentum number, and therefore, we expect it to significantly
modify the Hawking radiation spectrum. The greybody factor can be computed by determining
the absorption probability, $|{\cal A}_{j,n}^{(s)}|^2$, for propagation in the aforementioned
higher-dimensional background (\ref{bhmetric}), and then using the generalised
$(4+n)$-dimensional optical theorem relation~\cite{GKT}
\begin{equation}
\sigma^{(s)}_{j,n} (\omega) = \frac{2^{n}\pi^{(n+1)/2}\,\Gamma[(n+1)/2]}
{ n!\,\omega^{n+2}}\,\frac{(2j+n+1)\,(j+n)!}{j !}\,|{\cal A}^{(s)}_{j,n}|^2\,.
\label{grey-n}
\end{equation}
We should finally stress here the extremely important fact that the greybody factor
depends also on the dimensionality of spacetime, and as we will see, in the case of
a non-asymptotically flat spacetime, on the value of the bulk cosmological constant,
too. Therefore, the corresponding black hole radiation spectrum encodes valuable
information for the structure of the spacetime around the black hole, including the
number of dimensions and the curvature of the spacetime in which the black hole lives.
It is this double-fold dependence of the black hole radiation spectrum that we are
planning to investigate in this work.


\section{Hawking radiation from a $(4+n)$-Dimensional SdS Black Hole on the Brane}

In this section, we will concentrate on the emission of Hawking radiation on the brane
from a decaying higher-dimensional Schwarzschild-de-Sitter black hole. We will
first present the line-element describing the geometry of the induced on the brane
gravitational background, and then write the general equation for the propagation of
a field with arbitrary spin $s$ in this geometry. By solving this equation both
analytically (under certain approximations) and numerically, the greybody factor for
scalar emission on the brane from the black hole will be determined.

\subsection{General equation for spin-$s$ particles on the brane}

Having described the higher-dimensional spacetime in Section 2, we now turn our
attention to the induced geometry on the 4-dimensional brane. This simply follows
by fixing the
values of the `extra' angular azimuthal coordinates, i.e. $\theta_i=\pi/2$, for
$i=2, ..., n+1$. This leads to the projection of the higher-dimensional line-element
(\ref{bhmetric}) on a 4-dimensional slice that plays the role of our 4-dimensional
world. The projected line-element has the form
\beq
ds^2 = - h(r)\,dt^2 + \frac{dr^2}{h(r)} + r^2\,(d\theta^2 + \sin^2\theta\,
d\varphi^2)\,,
\label{brane}
\eeq
where the subscript `1' from the remaining azimuthal coordinate has been dropped.
Note that the metric function $h(r)$ remains unchanged during the projection and
is still given by Eq. (\ref{h-fun}), therefore, its profile along the $r$-coordinate
-- that now has components only along the 3 non-compact spatial dimensions -- remains
the same.

All particles restricted to live on the 4-dimensional brane propagate in the
projected background (\ref{brane}). The equations of motion, for particles with
arbitrary spin $s$, in a curved background can be derived by using the Newman-Penrose
formalism \cite{NP,Chandra}. Under the assumption of minimal coupling with gravity,
and by employing the factorized ansatz
\begin{equation}
\Psi_s(t,r,\theta,\varphi)= e^{-i\omega t}\,e^{i m \varphi}\,\Delta^{-s}\, P_{s}(r)
\,S^{m}_{s,j}(\theta)\,,
\label{facto}
\end{equation}
the free equations of motion for particles with spin $s=0,\frac{1}{2}$ and 1 may
be combined to form a `master' equation, satisfied by the radial part of the field,
$P_s(r)$. A master equation for propagation in the projected on the brane background
of a $(4+n)$-dimensional rotating black hole was derived in \cite{Kanti} (see also
\cite{IOP}). In the limit of zero black hole angular momentum, the line-element
used in \cite{Kanti} reduces exactly to the spherically-symmetric one given in
Eq. (\ref{brane}). In that case, the master equation takes the form
\beq
\Delta^{s}\,\frac{d \,}{dr}\,\biggl(\Delta^{1-s}\,\frac{d P_s}{dr}\,\biggr) +
\biggl(\frac{\om^2 r^2}{h} + 2i s\,\om\,r -\frac{i s \om\,r^2 h'}{h}
- \tilde\lambda \biggr)\,P_s=0\,,
\label{master1}
\eeq
where $\Delta \equiv h r^2$, and $\tilde\lambda=\lambda +2 s$. The eigenvalue
$\lambda$ is defined through the angular `master' equation, satisfied by
$S^{m}_{s,j}(\theta)$ -- the so-called spin-weighted spherical harmonics
\cite{goldberg}; this has the form \cite{Kanti}
\beq
\frac{1}{\sin\theta}\,\frac{d \,}{d \theta}\,\biggl(\sin\theta\,
\frac{d S^m_{s,j}}{d \theta}\,\biggr) + \biggl[-\frac{2 m s \cot\theta}
{\sin\theta} - \frac{m^2}{\sin^2\theta} + s - s^2 \cot^2\theta 
+ \lambda \biggr]\,S^m_{s,j}=0\,.
\label{angular}
\eeq
It was found that $\lambda \equiv j\,(j+1)-s\,(s+1)$, and thus 
$\tilde\lambda=j\,(j+1)-s\,(s-1)$. We should note here that $\omega$ is the
energy of the propagating particle, $s$ its spin and $j$ its angular momentum
quantum number. 

Equation (\ref{master1}) was used in Refs. \cite{kmr1,kmr2,HK} to study the emission
of scalars, fermions, and gauge bosons by a ($4+n$)-dimensional Schwarzschild black
hole on the brane.
In the presence of a cosmological constant in the bulk, the exact expression of the
metric function $h(r)$ changes and assumes the form (\ref{h-fun}), but both projected
line-elements have the same general structure given in Eq. (\ref{brane}). Therefore,
Eq. (\ref{master1}) can also be used to study propagation of fields with arbitrary
spin $s$ in the projected  background of a higher-dimensional Schwarzschild-de-Sitter
black hole, and thus the emission of Hawking radiation from such a black hole on the
brane. This task, for the emission of scalar fields, will be performed in the next
two subsections: we will first derive analytically the low- and high-energy limit of
the greybody factor, and then we will proceed to derive exact numerical results for
the same quantity.

\subsection{Analytical results for scalar emission on the brane}

To determine the greybody factor $\sigma_{j,n}^{(s)}(\omega)$, one needs to find first
the absorption coefficient ${\cal A} _{j,n}^{(s)}(\omega)$ for
propagation of a field in the projected background (\ref{brane}). To this end, we
need to solve the radial equation (\ref{master1}) over the whole radial regime.
Even in the absence of a bulk cosmological constant, the general solution of this
equation is extremely difficult to be found. In Refs. \cite{kmr1,kmr2}, where the
emission of Hawking radiation from a higher-dimensional Schwarzschild black hole
was studied, this equation was solved analytically at the low-energy regime, 
i.e. for $\omega r_H \ll 1$, by making use of a well-known approximation method:
the solutions in the near-horizon and far-field regimes were found and then smoothly
matched in the intermediate zone. This led to an analytical expression for the
absorption coefficient as a function of the energy $\omega$, the spin $s$, the total
angular momentum number $j$, and the number of extra dimensions $n$ \cite{kmr1,kmr2}. 

In the presence of a cosmological constant in the bulk though, even the use of this
approximate method becomes extremely complicated, and the derivation
of a general analytical formula for the absorption coefficient is highly non-trivial.
Nevertheless, by using an alternative method, a less general but still analytical
expression for the absorption coefficient at the low-energy regime can be derived.
The method amounts to solving the radial equation in an intermediate zone, away from
the two horizons $r_H$ and $r_C$, in the infrared limit of $\omega~\rightarrow~0$.
This solution is then stretched towards the two horizons, and matched with the two
asymptotic solutions there. An analytical formula for the absorption coefficient
can be easily derived by looking at the $\ell=0$ partial wave, that, as we will see,
gives the dominant contribution to the absorption coefficient. This method was used
successfully in 4 dimensions ($n=0$) to derive a simplified analytical expression for
the absorption coefficient in the case of a Reissner-Nordstrom \cite{Gurcel} and
Schwarzschild-de-Sitter \cite{Brady} spacetime.

To our knowledge, a similar analysis for the case of a higher-dimensional SdS spacetime
projected onto a 4-dimensional brane has never been performed up to now. In what follows
we will address this question and concentrate on the case of scalar fields --
we will thus drop the spin index $s=0$ from the various quantities and replace $j$ by
$\ell$, the orbital angular momentum quantum number. Then, the radial equation
(\ref{master1}) will take the simpler form
\beq
\frac{d \,}{dr}\,\biggl(h\,r^2\,\frac{d P}{dr}\,\biggr) +
\biggl[\,\frac{\om^2 r^2}{h} - \ell (\ell+1)\,\biggr]\,P=0\,.
\label{scalar}
\eeq
In the intermediate zone away from the two horizons, and in the infrared limit of 
$\omega~\rightarrow~0$, the first term inside the square brackets can be ignored. The
remaining equation can then be integrated to derive the low-energy solution for $P(r)$,
for arbitrary $\ell$. In order to simplify our calculations, we will concentrate
directly on the mode $\ell=0$. Then, by using the definition of $h(r)$, Eq. (\ref{h-fun}),
the solution of the above equation, for an arbitrary number of transverse dimensions
$n$, takes the form
\beq
P(r) = C_1\,\Biggl[\frac{\log (r-r_H)}{2k_H r_H^2} -\frac{\log (r_C-r)}{2k_C r_C^2}
+ \sum_{m=1}^{n+1}\frac{\log (r+r_m)}{2k_m r_m^2}\Biggr]
+C_2\,. \label{inter}
\eeq
In the above expression, $C_1$ and $C_2$ are integration constants to be determined
shortly, and the $r_m$'s stand for the $n+1$ negative roots of the equation
$h(r)=0$. Finally, the quantity
\beq
k_i=\frac{1}{2}\,\frac{dh}{dr}\biggr|_{r=r_i}=\frac{1}{2 r_i}\,
\Bigl[\,(n+1) - \frac{2 \kappa^2_D \Lambda}{n+2}\,r_i^2\,\Bigr]\,, \label{surface}
\eeq
is the surface gravity at the location of the $i$th root, including $r_H$ and $r_C$. 
As it is clear from Eq. (\ref{inter}), in the limits $r \rightarrow r_H$
and $r \rightarrow r_C$, the first and second term, respectively, will be the
dominant one in each case with the remaining ones acquiring a constant value.

The solution near the two horizons can be alternatively found by solving the radial
equation (\ref{scalar}) directly in the limits $r \rightarrow r_H$ and $r \rightarrow
r_C$, a task that is greatly facilitated if we make use of the so-called `tortoise'
coordinate, defined by the relation
\beq
\frac{dr_*}{dr}=\frac{1}{h(r)}\,. \label{tortoise}
\eeq
Defining also a new radial function through the relation $u(r)=r P(r)$, Eq. (\ref{scalar})
takes the Schr\"odinger-like form
\beq
-\frac{d^2 u}{dr_*^2} + h(r)\biggl[\frac{\ell (\ell+1)}{r^2}+
\frac{h'(r)}{r}\,\biggr] u= \omega^2 u\,,
\eeq
or, more explicitly, 
\beq
-\frac{d^2 u}{dr_*^2} + h(r)\biggl[\frac{\ell (\ell+1)}{r^2}+
\frac{\mu\,(n+1)}{r^{n+3}} -\frac{4 \kappa^2_D \Lambda}{(n+2) (n+3)}
\,\biggr] u= \omega^2 u\,.
\label{eff-eq-br}
\eeq

From the above equation, we may easily read the gravitational potential barrier $V(r)$
that a scalar particle sees while propagating from the cosmological to the black hole
horizon or vice versa. As an illuminating example, Fig.~1 depicts the form of this barrier
for $n=1$, $\mu=1$, $\Lambda r_H^2=10^{-2}$ (henceforth, the values of the latter two
quantities will be given in terms of Planck units, i.e. $M_*^{n+1}$ and $M_*^{2+n}$,
respectively) and $\ell=0,1$ and 2.
We notice that the barrier takes its lowest possible value for the partial wave
$\ell=0$, therefore, it is this mode that is more likely to be emitted by the black hole,
with the emission of higher partial waves being considerably suppressed -- our decision,
therefore, to consider only the lowest partial wave in our analytical approach is
justified. Since the metric function $h(r)$ vanishes at both horizons, $r_H$ and $r_C$,
so does the potential barrier that is proportional to the metric function. Therefore,
in both of these regimes, the solution for $u$ will have the form of plane waves.
We may therefore write
\bea
u(r_*) &\simeq & A_1\,e^{-i \omega r_*}\,, \quad {\rm for} \quad r\simeq r_H\,,
\label{BH}\\[3mm]
u(r_*) &\simeq& B_1\,e^{-i \omega r_*} + B_2\,e^{i \omega r_*}\,, \quad
{\rm for} \quad r\simeq r_C\,. \label{CH}
\eea
The solution near the black hole horizon, Eq. (\ref{BH}) has been written in such a way
as to satisfy the boundary condition that only incoming modes must exist
in this area -- no such condition exists for the solution at the cosmological horizon,
which therefore includes both incoming and outgoing modes.

\begin{center}
\begin{figure}[t]
\hspace*{3cm}\includegraphics[scale=0.70]{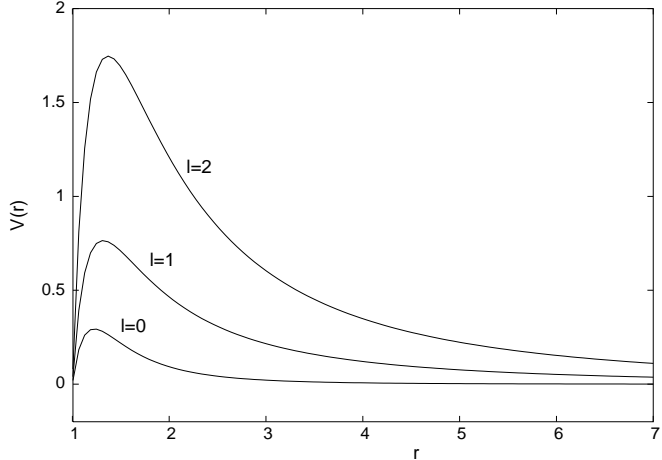}
\caption{\it The gravitational potential barrier $V(r)$ in the projected 4-dimensional
SdS spacetime for $n=1$, $\mu=1$, $\Lambda r_H^2=10^{-2}$ and $\ell =0,1$ and $2$.
The barrier vanishes at $r_H=1.0008$ and $r_C=24.474$.}
\end{figure}
\end{center}

\vspace*{-1.3cm}
By using the definition (\ref{h-fun}), Eq. (\ref{tortoise}) can be integrated
to give
\beq
r_*=
\frac{\log (r-r_H)}{2k_H} -\frac{\log (r_C-r)}{2k_C} +
\sum_{m=1}^{n+1}\frac{\log (r+r_m)}{2k_m}\,.
\eeq
In the low-energy limit, the asymptotic solution near the black hole horizon
($r \rightarrow r_H$), Eq. (\ref{BH}), can be written as
\bea
u^{(B)}(r) &\simeq& A_1\,\Biggl\{1-i\omega\,\biggl[
\frac{\log (r-r_H)}{2k_H} -\frac{\log (r_C-r_H)}{2k_C} + 
\sum_{m=1}^{n+1}\frac{\log (r_H+r_m)}{2k_m}\biggr]\Biggl\}\,.
\label{infra-BH}
\eea
Comparing the above with the function $r P(r)$, taken in the limit
$r \rightarrow r_H$, we find that
\beq
C_1=-i \omega r_H A_1\,, \qquad  C_2=\frac{A_1}{r_H} + {\cal O}(\omega)\,.
\label{ident1}
\eeq

A similar low-energy expansion of the asymptotic solution (\ref{CH}), 
near the cosmological horizon $r_C$, yields the following expression
\beq
u^{(C)}(r) \simeq (B_1+B_2) + i\omega\,(-B_1+B_2)\,
\biggl[\frac{\log (r_C-r_H)}{2k_H} -\frac{\log (r_C-r)}{2k_C} + 
\sum_{m=1}^{n+1}\frac{\log (r_C+r_m)}{2k_m}\biggr]\Biggl\}\,.
\label{infra-CH}
\eeq
Comparing again the expressions of $r P(r)$, in the limit $r \rightarrow r_C$,
and $u^{(C)}(r)$, we are led to the relations
\beq
C_1=i \omega r_C\,(-B_1+B_2)\,, \qquad  C_2=\frac{1}{r_C}\,(B_1+B_2) + 
{\cal O}(\omega)\,.
\label{ident2}
\eeq

As it is clear from the asymptotic solutions (\ref{BH})-(\ref{CH}), the reflection
coefficient for the propagation of a scalar field in this background will be given
by the ratio $B_2/B_1$, that can easily follow by combining Eqs. (\ref{ident1}) and
(\ref{ident2}). Then, the corresponding reflection and absorption probabilities
are found to be
\beq
|{\cal R}_{0,n}|^2=\Biggl|\frac{B_2}{B_1}\Biggr|^2=\frac{(r_C^2-r_H^2)^2}
{(r_C^2+r_H^2)^2}\,,
\qquad |{\cal A}_{0,n}|^2=1-|{\cal R}_{0,n}|^2=\frac{4r_C^2r_H^2}
{(r_C^2+r_H^2)^2}\,.
\label{reflection-b}
\eeq
The above results must be contrasted with the ones that follow in the case of  
propagation of a scalar field in the projected background of a Schwarzschild-like
higher-dimensional black hole. A detailed analytical calculation in Ref. \cite{kmr1}
showed that, in the absence of a bulk cosmological constant, the absorption
coefficient is given by: ${\cal A}_{\ell,n} \propto (\omega r_H)^{\ell+1}$, and therefore
vanishes in  the low-energy limit for all values of $\ell$. On the other hand, similar
results to the ones shown in Eqs. (\ref{reflection-b})
were derived also in the case of a scalar field being scattered in the interior
of a Reissner-Nordstrom black hole (i.e. between the Cauchy and the event horizon)
\cite{Gurcel}, and in the area between the black hole and cosmological horizon of
a 4-dimensional Schwarzschild-de-Sitter black hole \cite{Brady}. The deciding
factor therefore, for the value of the absorption coefficient in the infrared limit,
is the presence or not of a second horizon that creates a finite-size universe 
in which the particle propagates -- a particle with zero energy, i.e. infinite
wavelength, cannot be localized, and thus has a finite probability to traverse the
finite distance between the two horizons. As it was shown in Ref. \cite{Brady}, constant
field configurations, that have a vanishing energy-momentum tensor and thus zero
energy, do exist in a SdS background, therefore, we expect an enhancement in the
emission of ultra-soft quanta compared to the case of an asymptotically-flat
spacetime. We notice that, in the limit $\Lambda \rightarrow 0$, or $r_C \rightarrow
\infty$, the absorption coefficient vanishes, as expected for the scattering of a
zero-energy scalar particle in a single-horizon area.

Although the functional dependence of the absorption probability on the two horizons,
$r_H$ and $r_C$, is the same as in the case of a pure
4-dimensional SdS spacetime, our result has an implicit dependence on the value $n$ of
the transverse spacelike dimensions as well as on the value of the bulk cosmological
constant. The dependence on both of these parameters is depicted on the entries of
Table 1, for $\mu=1$. For fixed $\Lambda$, the absorption probability reduces as the
number of extra dimensions increases, a behaviour that was also encountered in the case of
propagation of a scalar field in a projected Schwarzschild-like black hole background
on the brane. On the other hand, for fixed $n$, the presence of an increasingly larger
bulk cosmological constant enhances the value of the absorption probability at the
low-energy regime, and thus
facilitates the emission of scalar fields from the black hole on the brane. The
dependence of the absorption probability on $n$ and $\Lambda$ could have also been
inferred from the qualitative dependence of the effective potential $V(r)$ on these
two parameters: keeping $\ell$ fixed, the height of the barrier increases as $n$
increases, while it decreases as $\Lambda$ increases.

\newpage

\medskip
\begin{center}
{\bf Table 1:} Absorption probabilities for emission on the brane in the limit
$\omega \rightarrow 0$\\[3mm]
$\begin{array}{|c|c||l|c|} \hline \hline
{\rule[-3mm]{0mm}{8mm}
\hspace*{0.6cm}{\bf n}\hspace*{0.6cm}} & \hspace*{0.2cm}{\bf |{\cal A}_{0,n}|^2}
\hspace*{0.2cm}({\rm for}\,\,\,\Lambda r_H^2=10^{-2})\hspace*{0.2cm} & 
\hspace*{0.5cm}{\bf \Lambda r_H^2} \hspace*{0.5cm} & \hspace*{0.2cm}
{\bf |{\cal A}_{0,n}|^2} \hspace*{0.2cm} ({\rm for}\,\,\,n=1) \hspace*{0.2cm} \\ \hline
{\rule[-1mm]{0mm}{6mm} 0} & 0.01417 & \hspace*{0.5cm} 0 & 0 \\
{\rule[-1mm]{0mm}{5mm} 1} & 0.00667 & \hspace*{0.3cm}10^{-4} & 6.67 \times 10^{-5} \\
{\rule[-1mm]{0mm}{5mm} 2} & 0.00340 & \hspace*{0.3cm}10^{-3} & 6.76 \times 10^{-4} \\ 
{\rule[-2mm]{0mm}{5mm} 3} & 0.00266 & \hspace*{0.3cm}10^{-2} & 6.67 \times 10^{-3} \\
{\rule[-2mm]{0mm}{5mm} 5} & 0.00142 & \hspace*{0.3cm}10^{-1.5} & 2.11 \times 10^{-2} \\
{\rule[-2mm]{0mm}{5mm} 7} & 0.00088 & \hspace*{0.3cm}10^{-1} & 6.67 \times 10^{-2} \\
\hline \hline 
\end{array}$
\end{center}
\bigskip \medskip

Let us now turn to the high-energy regime, where the greybody factor assumes its
geometrical optics limit value. This limiting value has been successfully used in
4-dimensional black hole backgrounds to describe the greybody factor of the
corresponding black hole \cite{page,sanchez,MTW}. This technique makes use
of the geometry around the emitting (or absorbing) body and can be clearly used
to perform the same calculation in more complicated gravitational backgrounds.
In particular, the same principle was used to derive the greybody factor, in the
high-energy regime, of a higher-dimensional Schwarzschild black hole projected onto
the brane \cite{emparan,HK}. We will apply here the same method for a SdS projected
black hole: for a massless particle in a circular orbit around a black hole, described
by  the line-element (\ref{brane}), we write its equation of motion, $p^\mu p_\mu=0$,
in the form
\begin{equation}
\biggl(\frac{1}{r}\,\frac{dr}{d\varphi}\biggr)^2=\frac{1}{b^2}
-\frac{h(r)}{r^2}\,,
\label{circular}
\end{equation}
where $b$ is the ratio of the angular momentum of the particle over its linear
momentum. Clearly, the classically accessible regime is defined by the relation
$b< {\rm min}(r/\sqrt{h})$. The structure of the particular gravitational background
enters in the above relation through the metric function $h(r)$. Using the definition
(\ref{h-fun}) and minimising the function $r/\sqrt{h}$ ,
we find that the maximum value of $b$, i.e. the closest the particle can get to
the black hole before getting absorbed, is given by 
\begin{equation}
b_c = r_H\,\Biggl\{\,\biggl(\frac{n+1}{n+3}\biggr)\,\biggl(\frac{n+3}{2}\,\biggl[\,1- 
\frac{2 \kappa^2_D \Lambda r_H^2}{(n+2) (n+3)}\,\biggr]\biggr)^{-2/(n+1)} -
\frac{2 \kappa^2_D \Lambda r_H^2}{(n+2) (n+3)}\Biggr\}^{-1/2}\,.
\label{effective}
\end{equation}
Then, the corresponding area, $\sigma_g = \pi b_c^2$, defines the absorptive area of
the black hole at high energies and, thus, its greybody factor -- being a constant,
the emitting body is characterized at high energies by a blackbody radiation spectrum.
For $n=0$ and $\Lambda=0$, the above expression reduces to the four-dimensional one,
$\sigma_g= 27 \pi r_H^2/4$ \cite{sanchez,MTW}. For $\Lambda=0$ but $n \neq 0$, we
recover the expression for the greybody factor at high energies of a ($4+n$)-dimensional
Schwarzschild black hole projected onto a brane \cite{emparan,HK}. The general
expression (\ref{effective}) has a strong dependence on the values of both the bulk
cosmological constant and the number of extra dimensions. If we keep $\Lambda$ fixed
and vary $n$, we obtain a suppression of $\sigma_g$ as $n$ increases, similar to the
one found in the case of brane emission of scalar fields from a projected Schwarzschild
black hole \cite{emparan,HK}. On the other hand, if we fix $n$ and vary $\Lambda$, the
greybody factor is enhanced as $\Lambda$ increases. Therefore, at the high-energy
regime, the presence of a bulk cosmological constant and a number of
transverse-to-the-brane dimensions have an increasing and decreasing, respectively,
effect on the emission cross-section of scalar fields from a projected
SdS black hole. The entries of Table 2 show the dependence of the greybody factor at
high-energies for some indicative values of $n$ and $\Lambda$, and for $\mu=1$.

\medskip
\begin{center}
{\bf Table 2:} Greybody factors for emission on the brane in the high-energy limit\\[3mm]
$\begin{array}{|c|c||l|c|c|} \hline \hline
{\rule[-5mm]{0mm}{13mm} \hspace*{0.4cm}{\bf n}\hspace*{0.4cm}} &
\hspace*{0.1cm}{\bf \begin{tabular}{c}({\rm for}\,\,$\Lambda r_H^2=10^{-2}$)\\[1mm]
{\bf $\sigma_g\,\,(\pi r_H^2)$}\end{tabular}}\hspace*{0.1cm} & 
\hspace*{0.4cm}{\bf \Lambda r_H^2} \hspace*{0.3cm} & 
\hspace*{0.1cm} {\bf \begin{tabular}{c}({\rm for}\,\,$n=0$)\\[1mm]
{\bf $\sigma_g\,\,(\pi r_H^2)$}\end{tabular}} \hspace*{0.1cm} & \hspace*{0.1cm}
{\bf \begin{tabular}{c}({\rm for}\,\,$n=1$)\\[1mm]
{\bf $\sigma_g\,\,(\pi r_H^2)$}\end{tabular}}\hspace*{0.1cm}\\ \hline
{\rule[-1mm]{0mm}{6mm} 0} & 6.8591 & \hspace*{0.5cm} 0 & 6.7500 & 4.0000 \\
{\rule[-1mm]{0mm}{5mm} 1} & 4.0201 & \hspace*{0.3cm}10^{-4} & 6.7511 & 4.0002 \\ 
{\rule[-1mm]{0mm}{5mm} 2} & 3.0774 & \hspace*{0.3cm}10^{-3} & 6.7607 & 4.0020 \\ 
{\rule[-1mm]{0mm}{5mm} 3} & 2.6017 & \hspace*{0.3cm}10^{-2} & 6.8591 & 4.0201 \\
{\rule[-1mm]{0mm}{5mm} 5} & 2.1179 & \hspace*{0.3cm}10^{-1.5} & 7.1114 & 4.0645 \\
{\rule[-1mm]{0mm}{5mm} 7} & 1.8699 & \hspace*{0.3cm}10^{-1} & 8.0962 & 4.2131\\
\hline \hline 
\end{array}$
\end{center}

\medskip
\subsection{Numerical results for the greybody factor on the brane}

To accurately evaluate the greybody factor in a wide energy range, and not only
in the low and high-energy regimes, it is necessary to go through numerical
computations. For this purpose, we first need to solve the scalar equation (\ref{scalar})
for propagation on the brane, and then to extract the amplitudes of the 
\textit{incoming} and \textit{outgoing} waves at the two asymptotic regimes
close to the two horizons.  The integration starts at the horizon of the black hole,
where the `no-outgoing' wave boundary condition can be easily applied, and
is propagated until the cosmological event horizon. Finally, the amplitudes 
of the \textit{incoming} and \textit{outgoing} waves are extracted by fitting
the analytical asymptotic solutions to the numerical results, and the absorption
cross-section (or, greybody factor) $\sigma_{\ell,n}(\omega)$ is then calculated in
terms of the absorption probability $|{\cal A}_{\ell,n}|^2$. 

A number of approximations have to be made in our numerical computation. First, in
order to get rid of the apparent singularity at $r=r_H$, the boundary condition at
the black hole horizon is applied at $r=r_H+\varepsilon$, where $\varepsilon$ must
be checked to be small enough so that our results are stable. Furthermore, for each
energy point, the sum over the angular momenta $\ell$, which defines the greybody factor
$\sigma_{n}(\omega)$, must be truncated at rank $m$, where $m$ has to be determined
so that the error to the value of the absorption cross section from omitting the
modes with $\ell>m$ is smaller than the accuracy of our calculation and thus irrelevant.
	
For the needs of the numerical integration of Eq. (\ref{scalar}), the vanishing
\textit{outgoing} wave boundary condition (\ref{BH}), applied at $r=r_H+\varepsilon$,
can be alternatively written as:
\begin{eqnarray}
	P\left(r\right) & = & 1 \label{BH-new-1}\\[2mm]
	\frac{{d}P}{{d}r} & = & -\frac{i\omega}{h\left(r\right)}.
	\label{BH-new}
\end{eqnarray}
Such a choice not only ensures a vanishing \textit{outgoing} wave at the black hole
horizon but also an incoming flux, in the same regime, equal to $\left|P\right|^2=1$
-- as our analytical arguments showed, the value of the integration constant $A_1$
at the black hole horizon does not have an observable effect, and therefore it can be
safely set, for simplicity, equal to unity.

Starting from the above asymptotic solution at the black hole horizon, the
numerical integration is then propagated towards the cosmological horizon. As the
latter horizon is approached, the consistency of the numerical computation requires
to vary the integration step according to the blue-shift suffered by the wave function.
According to the analytical arguments of the previous subsection, in this asymptotic
spacetime region, the radial wave function is the superposition of \textit{incoming}
and \textit{outgoing} plane waves:
\begin{equation}
	P(r)=B_1\,e^{-i\omega{r_*}}+B_2\,e^{i\omega{r_*}}\,,
	\label{cosmo_plane_sol}
\end{equation}
where the tortoise coordinate $r_*$ is given by Eq. (\ref{tortoise}). The above
asymptotic solution is particularly convenient for small values of the bulk
cosmological constant, when $r_C \gg r_H$ and the two asymptotic regimes of the
black hole and cosmological horizon are far apart. For larger values of $\Lambda$
though, an alternative far-field asymptotic solution can be used instead: by
making the change of variable
\beq
r \rightarrow f(r)=h(r)\,\biggl[\,1-\frac{2\kappa^2_D\Lambda r^2}{(n+2) (n+3)}\,\biggr]^{-1}\,,
\label{f}
\eeq
and the field redefinition
$P(f)=f^\alpha\,(1-f)^\beta\,F(f)$, the scalar equation (\ref{scalar}) takes, near
the cosmological horizon, the form of a hypergeometric equation\,\footnote{
The analysis leading to the derivation of the hypergeometric equation and to the
asymptotic solution near the cosmological horizon is similar to the one followed
in Refs. \cite{kmr1, kmr2, Kanti}.}, that finally leads to the solution
\begin{eqnarray}
&& \hspace*{-1cm}P(f)=B_1\,f^{\alpha}\,(1-f)^\beta\,F(a,b,c;f)
\nonumber \\[1mm] && \hspace*{2cm} +\,
B_2\,f^{-\alpha}\,(1-f)^\beta\,F(a-c+1,b-c+1,2-c;f)\,.
\label{NH-gen}
\end{eqnarray}
In the above, the indices ($a, b,c$) and power coefficients ($\alpha, \beta$)
depend on the fundamental parameters of the problem, i.e. the energy $\omega$
of the particle, the angular momentum number $\ell$, the number of extra dimensions
$n$, the value of the cosmological constant $\Lambda$, and the cosmological
horizon $r_C$. In the limit $r \rightarrow r_C$, the aforementioned solution
takes the asymptotic form 
\begin{equation}
	P(r)=B_1\,e^{-i\frac{\omega\,r_C}{A_C}\log[f(r)]}
	+B_2\,e^{i\frac{\omega\,r_C}{A_C}\log[f(r)]}\,,
	\label{alter}
\end{equation}
where $A_C=\left(n+1\right)-\frac{2\kappa^2_D\Lambda}{(n+2)}\,r_C^2$. The above
asymptotic solution takes better into account the curvature of the spacetime due
to the presence of the bulk cosmological constant, and provides a better fit to
our numerical results. Furthermore, the extraction of the \textit{incoming} and 
\textit{outgoing} amplitudes is straightforward. The new variable $f(r)$ has 
a profile along the radial coordinate similar to the one of $h(r)$: it vanishes
at $r=r_H$ and $r=r_C$ while it reaches a maximum value at some point in between.
Near the cosmological horizon, therefore, $\log[f(r)]$ decreases as $r$ increases;
taking also into account that $A_C$ is always negative, we may easily conclude that
$B_1$ is the amplitude of the \textit{incoming} wave and $B_2$ the amplitude of the 
\textit{outgoing} wave with respect to the black hole horizon.

\begin{figure}[t]
	\begin{center}
	\mbox{\includegraphics[scale=0.44]{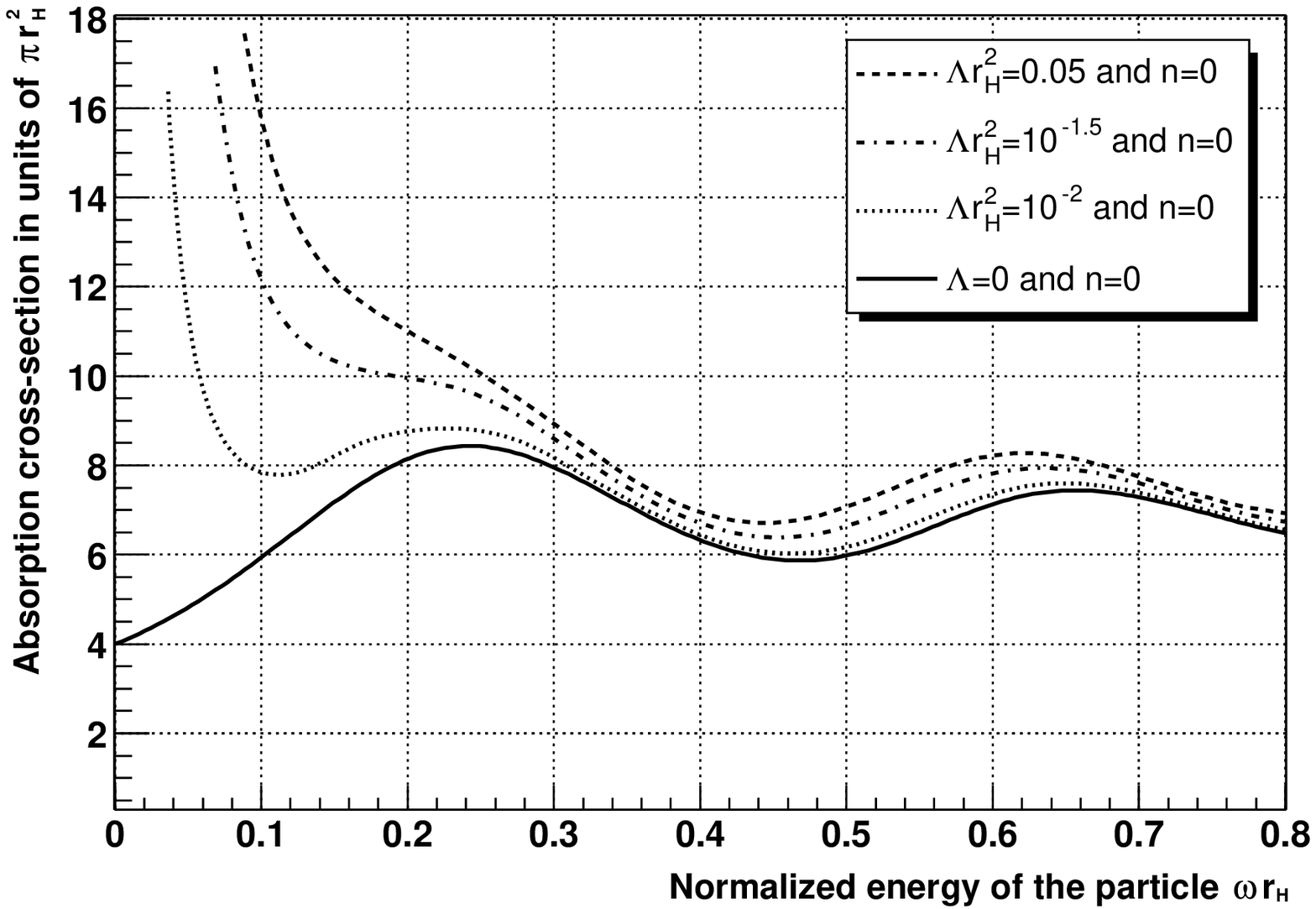}}
	{\includegraphics[scale=0.44]{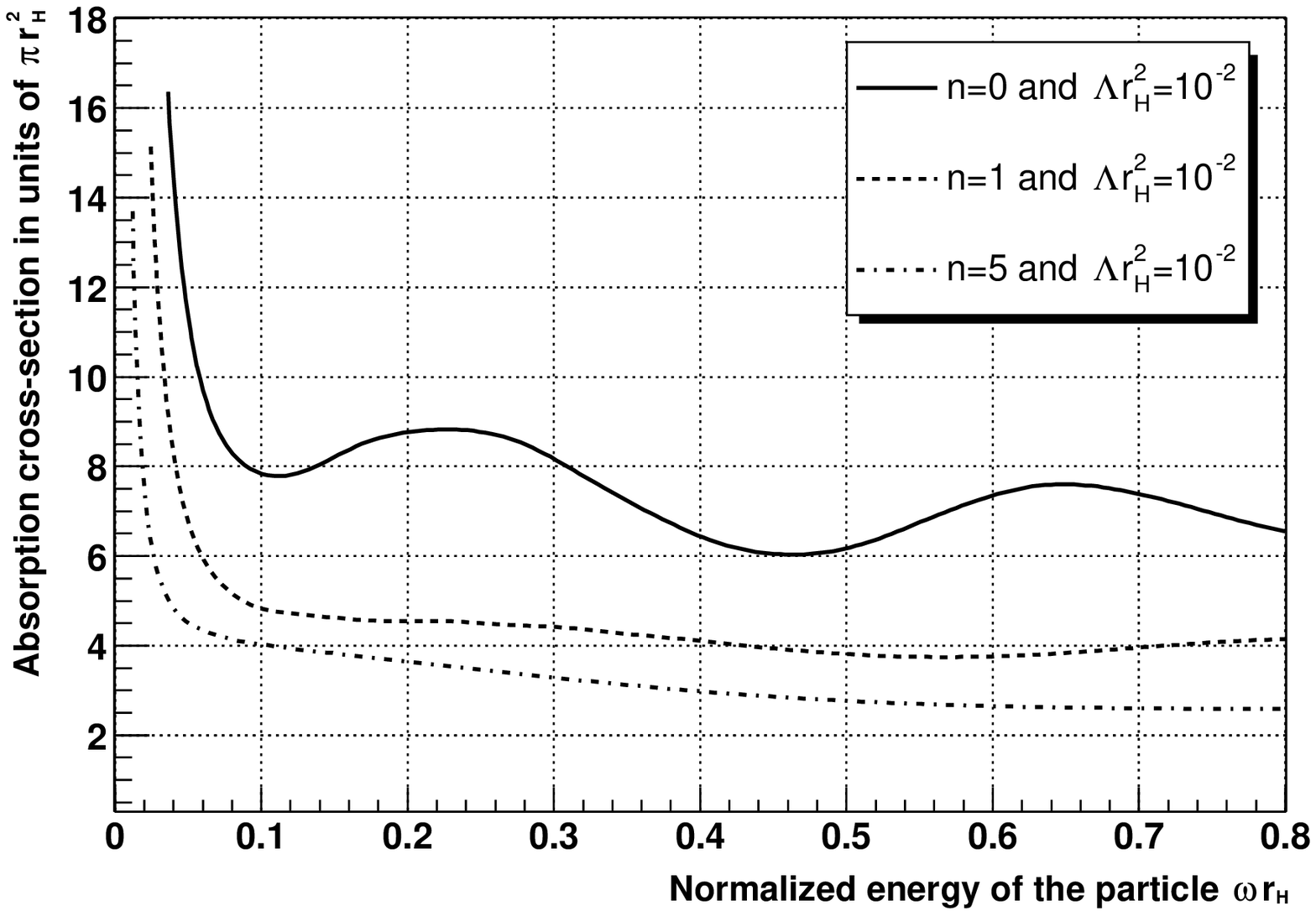}}
	\end{center} \vspace*{-5mm}
	\caption{ \it The absorption cross-section $\sigma_{n}(\omega)$, for scalar
	emission on the brane, versus the dimensionless parameter $\omega{r}_H$: {\rm (a)}
	in the first graph, the dimensionality of spacetime is fixed at $n=0$, while
	$\Lambda r_H^2$ takes the values $\{0,10^{-2},10^{-1.5},0.05\}$; {\rm (b)} in
	the second graph, the cosmological constant is fixed at $\Lambda r_H^2=10^{-2}$,
	while $n$ takes the indicative values $\{0,1,5\}$.}
	\label{res_sig_br}
\end{figure}


The absorption cross-section for the emission of fields on the brane follows from the
general expression (\ref{grey-n}) by setting $n=0$. In addition, the absorption
probability $|{\cal A}_{\ell,n}|^2$ can be directly determined from the ratio of the
incoming fluxes at the cosmological and black hole horizon \cite{Kanti}. We can 
then write the simple relation\,:
\begin{equation}
\sigma_{\ell,n}(\omega)=\frac{2\ell+1}{\omega^2}\,|{\cal A}_{\ell,n}|^2=
\frac{2\ell+1}{\omega^2}\,\left|\frac{1}{B_1}\right|^2\,.
\label{grey-4}
\end{equation}
Figures \ref{res_sig_br}(a,b) show the results obtained for $\sigma_{n}(\omega)$ for
scalar particles, after the summation over $\ell$ has been performed, and its dependence
on the value of the cosmological constant $\Lambda$ and the number of extra dimensions
$n$, respectively. In Fig. \ref{res_sig_br}a, the number of extra dimensions has been
kept fixed, at $n=0$, while the value of $\Lambda$ is allowed to vary. These numerical
results are in good agreement with the analytical ones, derived in the previous
subsection, both in the low and high-energy regime. When a positive cosmological
constant is introduced, the presence of the second (cosmological) horizon causes
the absorption probability  $|{\mathcal A}_{n}|^2$ to acquire a non-vanishing value
at low energies, which then causes a divergence in the value of $\sigma_n(\omega)$,
as $\omega\rightarrow0$, due to the $\omega^{-2}$ coefficient in Eq. (\ref{grey-4}). 
As a result, the well-known relation at the low-energy limit between the greybody
factor and the area of the horizon \cite{sanchez, kmr1, HK, Jung} ceases to
hold in the presence of a non-vanishing $\Lambda$.
In accordance to the entries of Tables 1 and 2, an increase in the value of the
cosmological constant causes an enhancement in the value of the greybody factor,
with the effect being more significant in the low-energy regime. In the high-energy
regime, the effect of the cosmological constant is less significant, nevertheless, 
as it can be seen from the above curves, the geometrical optics limit of the absorption 
cross-section does slightly increase as a function of $\Lambda$, in agreement with
the analytical results of the previous subsection.

The existence of additional dimensions in the universe also modifies the behaviour of the
absorption cross-section, as shown in Fig. \ref{res_sig_br}b. In the low energy regime,
an increase in the number of extra dimensions $n$, for a given value of $\Lambda$,
leads to a suppression of the value of the absorption cross-section. This can be
easily understood by noticing that increasing the number of dimensions $n$ leads to
an increase in the value of the cosmological horizon $r_C$, which is equivalent to
decreasing the value of $\Lambda$. In the high-energy limit, the effect of the
dimensionality of spacetime is the same as for asymptotically-flat spacetimes
\cite{HK}~: $\sigma_n(\omega)$ is suppressed as $n$ increases. The behaviour
obtained for the greybody factor as a function of $n$, both in the low and
high-energy regime, are in excellent agreement with the analytical results and the
entries of Tables 1 and 2 presented in section 3.2.

\section{Hawking radiation from a $(4+n)$-Dimensional SdS Black Hole in the Bulk}

According to the assumptions of the model, Standard Model fields (fermions,
gauge bosons, Higgs fields) are restricted to live on the 4-dimensional brane,
and therefore have no access to the transverse dimensions. Nevertheless,
scalar particles that carry no charges under the Standard Model gauge group
are in principle allowed to propagate in the bulk. Thus, in the study of the
process of the emission of Hawking radiation from a higher-dimensional
black hole into the bulk, scalar fields, apart from gravitons, should also
be considered. Although this part of radiation would clearly be invisible to
an observer on the brane, estimating the amount of energy lost in
the ``bulk" channel is imperative before accurate predictions for the emitted
radiation on the brane can be made. 

The field equation describing the motion of a scalar particle in a 
higher-dimensional background is easy to find. For a free scalar field
minimally coupled to gravity, its equation of motion takes the form
$D_M D^M \phi=0$, where $D_M$ is the covariant derivative in the
higher-dimensional spacetime. For a spherically-symmetric, $(4+n)$-dimensional
gravitational background of the form (\ref{bhmetric}), this equation takes a
separable form if we use the factorized ansatz
\beq
\phi(t,r,\theta_i,\varphi)=
e^{-i\om t}\,P(r)\,{\tilde Y}_\ell(\Om)\, ,
\eeq
where ${\tilde Y}_\ell(\Om)$ is the $(3+n)$-spatial-dimensional generalisation
of the usual spherical harmonic functions depending on the angular coordinates
\cite{Muller}. Upon using this ansatz, the scalar wave equation reduces to a
second-order differential equation for the radial wavefunction $P(r)$:
\beq
\frac{h(r)}{r^{n+2}}\,\frac{d \,}{dr}\,
\biggl[\,h(r)\,r^{n+2}\,\frac{d P}{dr}\,\biggr] +
\biggl[\,\om^2 - \frac{h(r)}{r^2}\,\ell\,(\ell+n+1)\,\biggr] P =0 \, .
\label{bulk-eqn}
\eeq
The explicit dependence of the above equation on the dimensionality of
spacetime is evident: it comes from the contribution of the $(n+2)$, in total,
compact dimensions, and from the eigenvalue of the $(3+n)$-dimensional
spherical harmonic functions. An additional implicit dependence enters
through the expression of the metric function $h(r)$. The latter is still given
by Eq. (\ref{h-fun}), and it is thus characterized, as before, by two positive
roots, $r_H$ and $r_C$, and $(n+1)$ negative ones collectively denoted by $r_m$.

The above radial equation has been studied before, both analytically
\cite{kmr1, Frolov1} and numerically \cite{HK}, in the absence of a bulk
cosmological constant, to derive the spectrum of Hawking radiation from
a higher-dimensional Schwarzschild-like black hole. Although the principles
behind the emission of Hawking radiation in the bulk and on the brane are
the same, the effect of the dimensionality of spacetime on the corresponding
greybody factors and final emission rates was found to be different in the two
``channels".
In the presence of a bulk cosmological constant, we expect the existence of
the second horizon $r_C$, to modify the low-energy bulk absorption coefficient
in a way similar to the one for brane emission. Nevertheless, we anticipate the
dimensionality of spacetime to have a much more explicit effect compared to the
case of brane emission, where the number of additional dimensions appeared
only implicitly in Eqs. (\ref{reflection-b}) through the values of $r_H$ and $r_C$.
In the next two subsections, we focus on the emission of scalar fields in the bulk,
and address the problem of finding the corresponding greybody factor both
analytically and numerically.

\subsection{Analytical results for scalar emission in the bulk}

In this subsection, we follow the same methods as the ones presented in section 3.2 
in order to derive analytically the low-energy absorption probability $|{\cal A}_{0,n}|^2$
and the high-energy absorption cross-section $\sigma_g$ for the propagation of scalar
fields in the background of the $(4+n)$-dimensional black hole given by
Eqs. (\ref{bhmetric}) and (\ref{h-fun}).

We will start by deriving the solution for the $s$-wave partial mode in the infrared
limit, and compare it with the asymptotic solutions in the vicinity of the two
horizons, $r_H$ and $r_C$, expanded in the low-energy limit. For propagation in
the bulk, the former can be derived by integrating the radial equation
(\ref{bulk-eqn}) in the intermediate radial regime ($r_H \ll r \ll r_C$), for
$\ell=0$ and $\omega \rightarrow 0$. We then find
\beq
P(r) = C_1\,\Biggl[-\frac{\log r}{\mu} + \frac{\log (r-r_H)}{2k_H\,r_H^{2+n}} 
-\frac{\log (r_C-r)}{2k_C\,r_C^{2+n}} + 
\sum_{m=1}^{n+1}\frac{\log (r+r_m)}{2k_m\,r_m^{2+n}}\biggr] +C_2\,.
\label{P-bulk}
\eeq
In the above, $k_i$ stands again for the surface gravity at the location of the
$i$th root of the equation $h(r)=0$, and it is still given by Eq. (\ref{surface}).

\begin{center}
\begin{figure}[t]
\hspace*{3cm}
\includegraphics[scale=0.7]{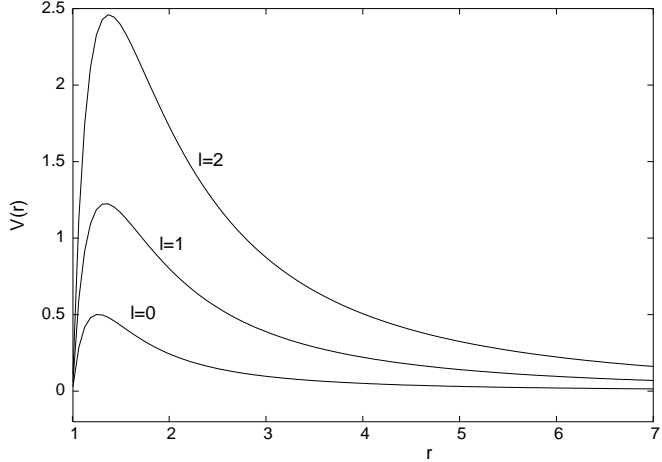}
\caption{\it The gravitational potential barrier $V(r)$ in the $(4+n)D$ SdS spacetime,
for $n=1$, $\mu=1$, $\Lambda r_H^2=10^{-2}$ and $\ell =0,1$ and $2$. The barrier
vanishes again at $r_H=1.0008$ and $r_C=24.474$ but its height is significantly higher
than on the brane.} \label{Veff-bulk}
\end{figure}
\end{center}

By using the definition $u(r)=r^{(n+2)/2}\,P(r)$, and the same tortoise coordinate
defined in Eq. (\ref{tortoise}), the scalar field equation in the bulk
(\ref{bulk-eqn}) takes the Schr\"odinger-like form 
\beq
-\frac{d^2 u}{dr_*^2} + h(r)\Biggl[\frac{\ell (\ell+n+1)}{r^2}+
\frac{(n+2)\,h'(r)}{2r} +\frac{n(n+2)\,h(r)}{4r^2}\,\Biggr] u= \omega^2 u\,,
\eeq
or, more explicitly,
\beq
-\frac{d^2 u}{dr_*^2} + h(r)\Biggl[\frac{(2\ell+n+1)^2-1}{4 r^2}+
\frac{\mu\,(n+2)^2}{4 r^{n+3}} -\frac{\kappa^2_D \Lambda\,(n+4)}{2\,(n+3)}\,\Biggr] u=
\omega^2 u\,. \label{eff-eq-bulk}
\eeq
Once again, from the above expression, one may easily read the gravitational
barrier $V(r)$ seen by a scalar field propagating in the higher-dimensional bulk. 
In Fig. \ref{Veff-bulk}, this gravitational barrier is shown for the indicative values
of $n=1$, $\mu=1$, $\Lambda r_H^2=10^{-2}$ and $\ell =0,1$ and $2$. As in the case of
propagation on the brane, the barrier increases for increasing $\ell$, therefore,
it is the $s$-wave that is more likely to be emitted also in the bulk. Nevertheless,
we notice that the height of the barrier is significantly higher compared to the one
for propagation on the brane, therefore, we expect the emissivity of bulk
scalar modes to be suppressed compared to the one of brane scalar modes, at least
in the low- and intermediate-energy regime.
\vspace*{-1.3cm}

The metric function $h(r)$ is common for propagation both on the brane and in
the bulk, therefore the potential vanishes at the same values of the radial
coordinate, i.e. at $r_H$ and $r_C$. In these asymptotic regimes, the solutions
are given again by Eqs. (\ref{BH})-(\ref{CH}), and their infrared limit by
Eqs. (\ref{infra-BH}) and (\ref{infra-CH}). Comparing them with the $s$-wave
solution (\ref{P-bulk}), stretched towards the two horizons, will allow us to
determine the relations between the set of coefficients $A_1$, ($C_1, C_2$)
and ($B_1, B_2$). In particular, comparing $u^{(B)}(r_*)$, Eq. (\ref{infra-BH}),
with $r^{(n+2)/2} P(r)$, with the latter taken in the limit $r \rightarrow r_H$,
we obtain the relations
\beq
C_1=-i \omega\,r_H^{(n+2)/2} A_1\,, \qquad  C_2=\frac{A_1}{r_H^{(n+2)/2}} + 
{\cal O}(\omega)\,.
\label{ident3}
\eeq
Similarly, comparing $u^{(C)}(r_*)$, Eq. (\ref{infra-CH}), with $r^{(n+2)/2} P(r)$,
in the limit $r \rightarrow r_C$, we find in turn
\beq
C_1=i \omega r_C^{(n+2)/2}\,(-B_1+B_2)\,, \qquad  
C_2=\frac{1}{r_C^{(n+2)/2}}\,(B_1+B_2) + 
{\cal O}(\omega)\,.
\label{ident4}
\eeq
Rearranging the above set of constraints, we may easily derive the ratio $B_2/B_1$.
Then, the reflection and absorption probabilities for the propagation of a scalar
field in the bulk are found to be
\beq
|{\cal R}_{0,n}|^2=\Biggl|\frac{B_2}{B_1}\Biggr|^2=\Biggl(\frac{r_C^{n+2}-r_H^{n+2}}
{r_C^{n+2}+r_H^{n+2}}\Biggr)^2\,,
\qquad |{\cal A}_{0,n}|^2=\frac{4\,(r_C\,r_H)^{n+2}}{(r_C^{n+2}+r_H^{n+2})^2}\,.
\label{reflection}
\eeq
As in the case of brane emission, the absorption probability for emission in the bulk
is not zero in the low-energy limit, contrary to what was derived for the propagation
of a scalar field in the asymptotically-flat, $(4+n)$-dimensional  background of a
Schwarzschild black hole, where $|{\cal A}_{\ell,n}|^2 \sim (\omega r_H)^{2\ell+n+2}$
\cite{kmr1}. The above result differs also from the one derived in section 3.2 for
brane propagation, Eq. (\ref{reflection-b}),
in that, as hinted at the beginning of section 4, the dimensionality of the
spacetime in which the field propagates appears explicitly in the expression of 
the absorption probability at low energies. Table 3 shows the exact dependence of
the absorption probability, for emission in the bulk at the low-energy limit, on the
number of extra dimensions $n$ and the bulk cosmological constant $\Lambda$, for $\mu=1$.
Similarly to the case of emission on the brane, the presence of a bulk cosmological
constant enhances the absorption probability at low-energies and, thus, the emissivity
of the black hole into the bulk, while a number of extra dimensions present, $n>0$,
have a suppressing effect on it. A simple comparison of the entries of Tables 1 and 3
confirms our expectation of a significantly reduced emission in the bulk of the
dominant $s$-wave, compared to the one on the brane, for the same values of the free
parameters of the theory.

\begin{center}
{\bf Table 3:} Absorption probabilities for emission in the bulk in the limit
$\omega \rightarrow 0$\\[3mm]
$\begin{array}{|c|c||l|c|} \hline \hline
{\rule[-3mm]{0mm}{8mm}
\hspace*{0.6cm}{\bf n}\hspace*{0.6cm}} & \hspace*{0.2cm}{\bf |{\cal A}_{0,n}|^2}
\hspace*{0.2cm}({\rm for}\,\,\,\Lambda r_H^2=10^{-2})\hspace*{0.2cm} & 
\hspace*{0.5cm}{\bf \Lambda r_H^2} \hspace*{0.5cm} & \hspace*{0.2cm}
{\bf |{\cal A}_{0,n}|^2} \hspace*{0.2cm} ({\rm for}\,\,\,n=1) \hspace*{0.2cm} \\ \hline
{\rule[-1mm]{0mm}{6mm} 0} & 1.417 \times 10^{-2}& \hspace*{0.5cm} 0 & 0 \\
{\rule[-1mm]{0mm}{5mm} 1} & 2.735 \times 10^{-4}& \hspace*{0.3cm}10^{-4} & 
2.722 \times 10^{-7} \\
{\rule[-1mm]{0mm}{5mm} 2} & 4.006 \times 10^{-6} & \hspace*{0.3cm}10^{-3} & 
8.611 \times 10^{-6} \\ 
{\rule[-2mm]{0mm}{5mm} 3} & 4.594 \times 10^{-8} & \hspace*{0.3cm}10^{-2} & 
2.735 \times 10^{-4} \\
{\rule[-2mm]{0mm}{5mm} 5} & 3.445 \times 10^{-12} & \hspace*{0.3cm}10^{-1.5} & 
1.554 \times 10^{-3} \\
{\rule[-2mm]{0mm}{5mm} 7} & 1.455 \times 10^{-16} & \hspace*{0.3cm}10^{-1} & 
9.019 \times 10^{-3} \\
\hline \hline 
\end{array}$
\end{center}
\medskip

We then turn to the high-energy regime, and to the geometrical optics limit value
of the greybody factor. The equation describing the motion of a particle, with
only one non-vanishing component of angular momentum, in a circular orbit around
a black hole is given by Eq. (\ref{circular}), for propagation both in the bulk
and on the brane -- this equation is derived by fixing the values of all azimuthal
angular coordinates $\theta_i$ to $\pi/2$, therefore the existence or not of $n$
additional such variables does not change the final result. The classically accessible
regime is thus again given by $b< {\rm min}(r/\sqrt{h})$ with the minimum radial
proximity given by Eq. (\ref{effective}). What nevertheless changes, in the case
of propagation in a higher-dimensional black hole background, is the expression
of the absorptive area of the black hole as a function of the critical radius
$b_c$: as it was pointed out in Ref. \cite{HK}, when carefully computed, this
area comes out to be
\begin{equation}
A_p = \frac{2 \pi}{(n+2)}\,\frac{\pi^{n/2}}{\Gamma[(n+2)/2]}\,
b_c^{\,n+2}\,.
\end{equation}
This area stands for the absorption cross-section, or greybody factor, at high
energies. In the case of a higher-dimensional background with a bulk cosmological
constant, that we study here, it takes the explicit form
\bea
\sigma_{g} &=& 
\frac{A_H}{\sqrt{\pi}\,(n+2)}\,\frac{\Gamma[(n+3)/2]}{\Gamma[(n+2)/2]}\, \times
\\[3mm] & \times &  
\,\Biggl\{\,\biggl(\frac{n+1}{n+3}\biggr)\,\biggl(\frac{n+3}{2}\,\biggl[\,1- 
\frac{2 \kappa^2_D \Lambda r_H^2}{(n+2) (n+3)}\,\biggr]\biggr)^{-2/(n+1)} -
\frac{2 \kappa^2_D \Lambda r_H^2}{(n+2) (n+3)}\Biggr\}^{-(n+2)/2}
\hspace*{-1cm}\,, \nonumber 
\label{high}
\eea
upon making use of the expression for $b_c$ given in Eq. (\ref{effective}). In the
above, $A_H$ is the horizon area of the $(4+n)$-dimensional SdS black hole given by
\beq
A_H=r_H^{n+2} A_{2+n}=r_H^{n+2}\,(2 \pi)\,\pi^{(n+1)/2}\,\Gamma 
\biggl(\frac{n+3}{2}\biggr)^{-1}\,.
\eeq

In Table 4, we give some indicative values of the greybody factor, in units of the area
of the $(4+n)$-dimensional horizon $A_H$, for emission of scalar
fields in the bulk at high-energies, for particular values of $n$ and $\Lambda$, and for
$\mu=1$. Similarly to the case of emission on the brane, the presence of a bulk
cosmological constant enhances the greybody factor, and increases the emissivity of
the black hole into the bulk in the high-energy regime. In contrast to what happens
in the emission of brane modes, however, choosing larger values for the number of
transverse dimensions enhances the value of the greybody factor, instead of suppressing
it. The same effect takes place also in the case of emission of scalar fields in the
bulk by an asymptotically flat, higher-dimensional Schwarzschild black hole. The
above formula (\ref{high}) is a generalization of the one found in \cite{HK} for 
a non-vanishing bulk cosmological constant. Also in that simplified case, it may
be shown that, for increasingly larger $n$, the greybody factor is indeed enhanced at
high energies.
This enhancement in the emission of highly-energetic scalar fields in the bulk,
compared to the one on the brane, caused the boost in the bulk-to-brane emissivity for
large values of $n$ where the effect becomes significant \cite{HK}.

\newpage
\begin{center}
{\bf Table 4:} Greybody factors for emission in the bulk in the high-energy limit\\[3mm]
$\begin{array}{|c|c||l|c|c|} \hline \hline
{\rule[-5mm]{0mm}{13mm} \hspace*{0.4cm}{\bf n}\hspace*{0.4cm}} &
\hspace*{0.1cm}{\bf \begin{tabular}{c}({\rm for}\,\,$\Lambda r_H^2=10^{-2}$)\\[1mm]
{\bf $\sigma_g\,\,(A_H)$}\end{tabular}}\hspace*{0.1cm} & 
\hspace*{0.4cm}{\bf \Lambda r_H^2} \hspace*{0.3cm} & 
\hspace*{0.1cm} {\bf \begin{tabular}{c}({\rm for}\,\,$n=0$)\\[1mm]
{\bf $\sigma_g\,\,(A_H)$}\end{tabular}} \hspace*{0.1cm} & \hspace*{0.1cm}
{\bf \begin{tabular}{c}({\rm for}\,\,$n=1$)\\[1mm]
{\bf $\sigma_g\,\,(A_H)$}\end{tabular}}\hspace*{0.1cm}\\ \hline
{\rule[-1mm]{0mm}{6mm} 0} & 1.7148 & \hspace*{0.4cm} 0 & 1.6875 & 1.6977\\
{\rule[-1mm]{0mm}{5mm} 1} & 1.7105 & \hspace*{0.2cm}10^{-4} & 1.6878 & 1.6978 \\ 
{\rule[-1mm]{0mm}{5mm} 2} & 1.7757 & \hspace*{0.2cm}10^{-3} & 1.6902 & 1.6989 \\ 
{\rule[-1mm]{0mm}{5mm} 3} & 1.8535 & \hspace*{0.2cm}10^{-2} & 1.7148 & 1.7105 \\
{\rule[-1mm]{0mm}{5mm} 5} & 2.0117 & \hspace*{0.2cm}10^{-1.5} & 1.7778 & 1.7389 \\
{\rule[-1mm]{0mm}{5mm} 7} & 2.1622 & \hspace*{0.2cm}10^{-1} & 2.0241 & 1.8351\\
\hline \hline 
\end{array}$
\end{center}

\bigskip
\subsection{Numerical results for the greybody factor in the bulk}

As in the case of brane emission, in order to derive accurate results for the
emission of scalar fields in the bulk over the whole energy regime, we need to
perform a numerical computation. The method followed here is identical to the one
described in section 3.3. The numerical integration of the scalar bulk equation
(\ref{bulk-eqn}) starts from a point $r=r_H+ \varepsilon$, close to the black hole
horizon with the corresponding asymptotic solution satisfying the boundary conditions
(\ref{BH-new-1})-(\ref{BH-new}). The
solution is then propagated towards the cosmological horizon and the numerical
results are fitted by the appropriate asymptotic solution near $r_C$. As in
the case of brane emission, a better fit is provided by making use of the 
alternative asymptotic solution (\ref{alter}) -- although the scalar field
equation and its general solution in the bulk is different from the ones on the
brane, the asymptotic solution, for $r \rightarrow r_C$, has exactly the same form.
The absorption probability $|{\cal A}_{\ell,n}|^2$ can then be defined by the ratio
of the incoming fluxes at the black hole and cosmological horizon, and can
easily follow once the amplitude of the {\textit{ingoing}} wave $B_1$ is determined
from the fit of our numerical results. 

For emission in the bulk, the higher-dimensional formula, Eq. (\ref{grey-n}), for the
absorption cross-section $\sigma_{n}(\omega)$ must be used. We are interested both in
the effect of the cosmological constant and the dimensionality of spacetime on the
behaviour of the greybody factor, and these effects are depicted in Figs.
{\ref{res_sig_bulk}\,(a,b). In Fig. {\ref{res_sig_bulk}a, we have kept
the dimensionality of spacetime fixed, at $n=1$, and varied the value of $\Lambda$.
From the depicted curves, we
may easily conclude that the absorption cross-section is enhanced as the value
of $\Lambda$ increases: as in the case of brane emission, the effect is more 
significant at the low-energy regime and less significant at the high-energy
one, in complete accordance with the entries of Tables 3 and 4. For $\Lambda=0$,
we recover the well-known result of the greybody factor being equal to the 
horizon area of the higher-dimensional black hole \cite{kmr1, HK, Jung}.
Nevertheless, for non-vanishing values of $\Lambda$, the greybody factor
diverges in the low-energy limit due to the fact that the absorption
probability $|{\cal A}_{\ell,n}|^2$ acquires a non-vanishing constant value, 
in agreement to the analytical results of the previous subsection.

\begin{figure}[t]
	\begin{center}
	\mbox{\includegraphics[scale=0.44]{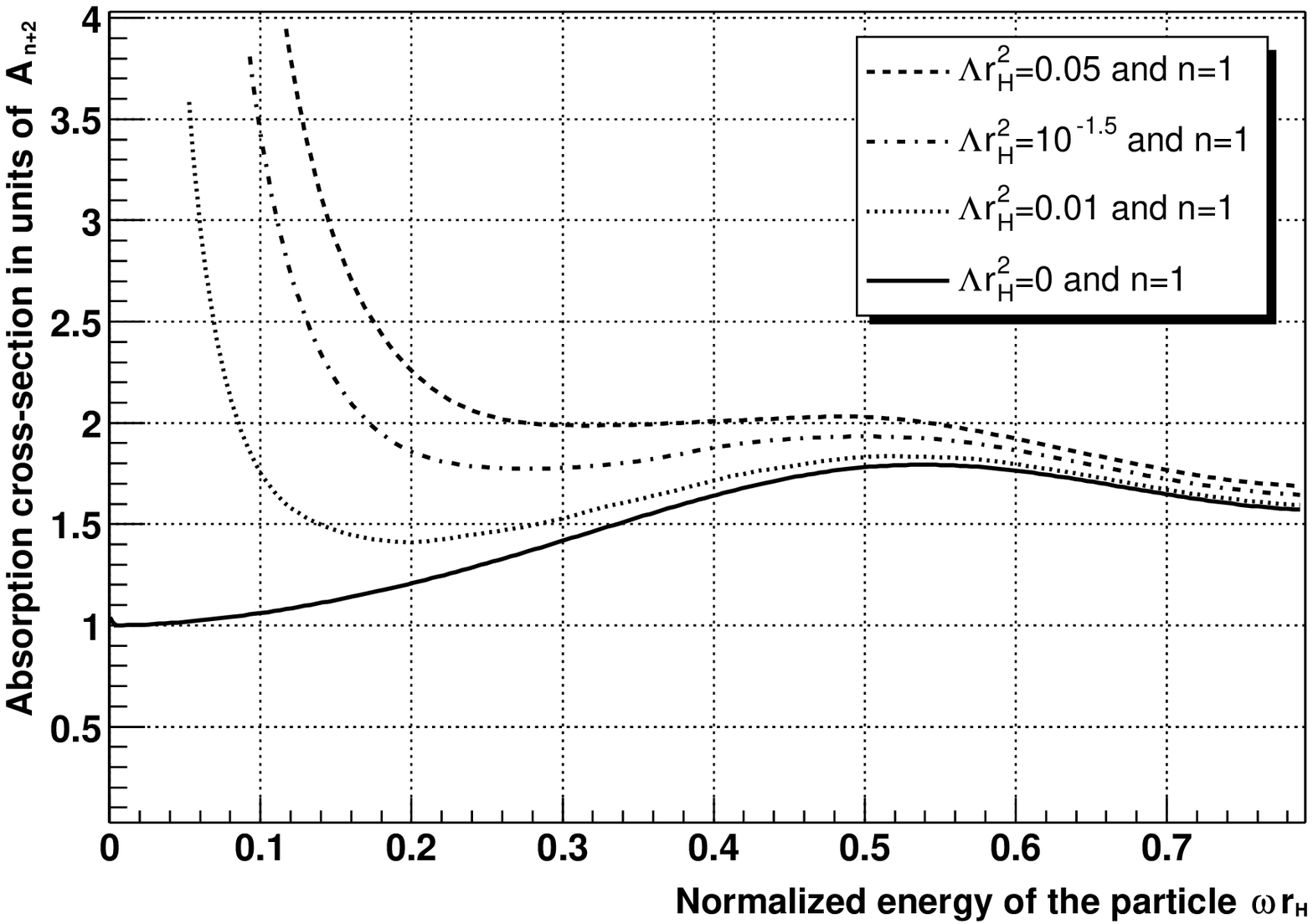}}
	{\includegraphics[scale=0.44]{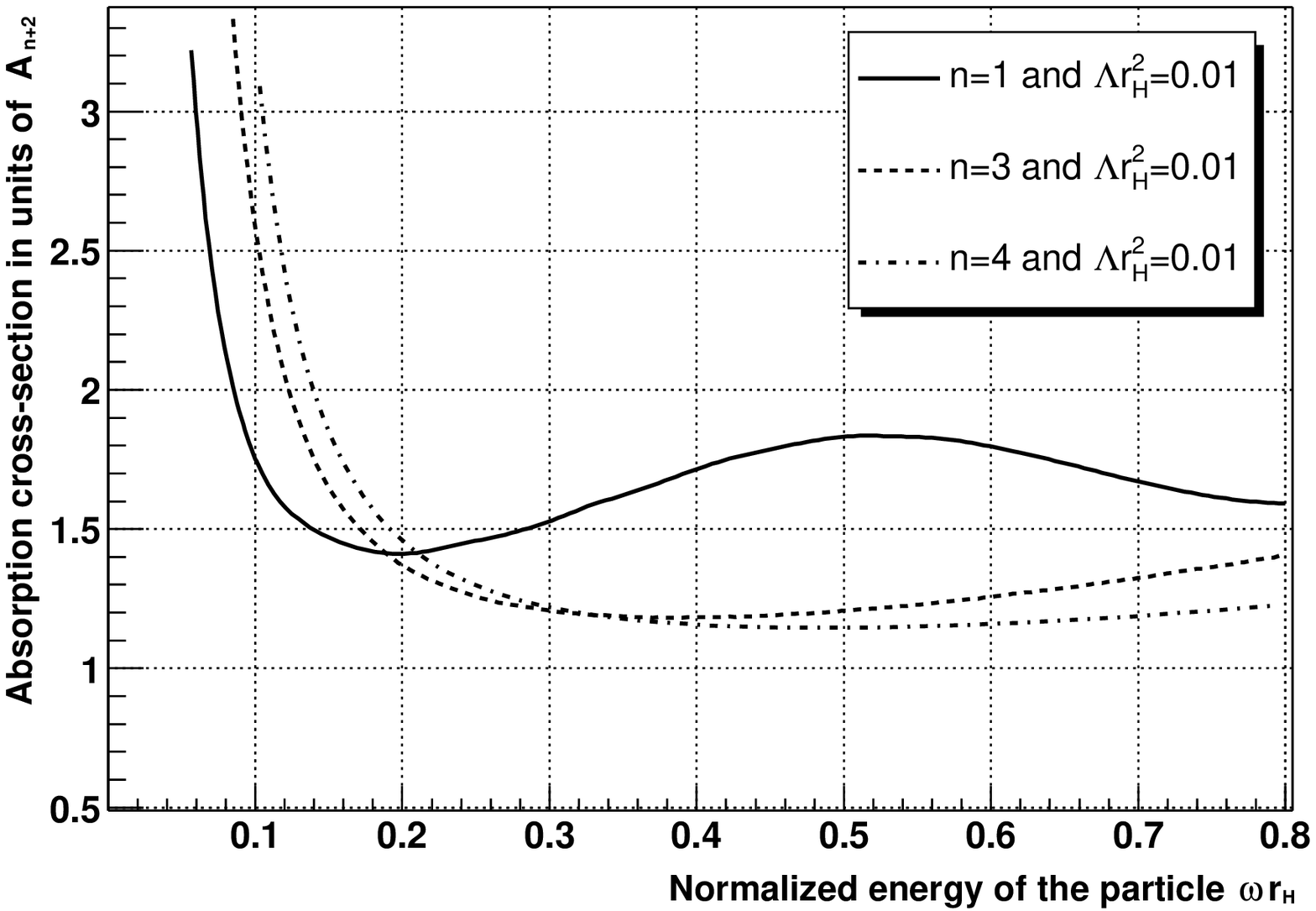}}
	\end{center} \vspace*{-5mm}
	\caption{ \it The absorption cross-section $\sigma_{n}(\omega)$, for scalar
	emission in the bulk, versus the dimensionless parameter $\omega{r}_H$:
	{\rm (a)} the dimensionality of spacetime is fixed at $n=1$, while $\Lambda r_H^2$
	takes the values $\{0,10^{-2},10^{-1.5},0.05\}$ in Planck units; {\rm (b)} the
	cosmological constant is fixed at $\Lambda r_H^2=10^{-2}$,
	and $n$ takes the indicative values $\{1,3,4\}$.}
\label{res_sig_bulk}
\end{figure}

In Fig. {\ref{res_sig_bulk}b, on the other hand, we have kept the value of the
cosmological constant fixed and varied the number of extra dimensions. The analytical
results for the absorption probability $|{\cal A}_{0,n}|^2$, derived in the previous subsection,
may have hinted towards the suppression of the greybody factor, too, in the low-energy
regime. Nevertheless, not only is the suppression of $|{\cal A}_{0,n}|^2$, as a function of $n$,
rather insignificant but also, as we may see from Eq. (\ref{grey-n}), there are actually
two sources of enhancement of $\sigma_{n}(\omega)$, as $n$ increases, in the
low-energy regime: the multiplicative, $n$-dependent coefficient, and, more importantly,
the factor $\omega^{-(n+2)}$. As a result, the greybody factor is indeed enhanced in
the low-energy regime, as the curves in Fig. \ref{res_sig_bulk}b also reveal. 
In the high-energy regime, finally, our curves seem to indicate that the greybody factor
decreases with $n$, instead of increasing, as our analytical results showed. This is
actually caused by the fact that, as in the case of emission in a flat spacetime \cite{HK},
the high-energy limit behaviour of the greybody factor in the bulk is attained for fairly
large values of $\omega r_H$; as a result, the curves in Fig. \ref{res_sig_bulk}b depict
its behaviour only up to intermediate energy values, where $\sigma_{n}(\omega)$ is indeed
suppressed with $n$. Extending our graph to higher values of energy leads to
agreement with our analytical results.


\section{Energy emission rates} 

In this section, we proceed to calculate the final energy emission rates for scalar
fields by a higher-dimensional SdS black hole, both in the bulk and on the brane, by
using the exact numerical results, derived in sections 3 and 4 for the corresponding
absorption probabilities/greybody factors. Once this task is performed, we compare the
amount of energy emitted in the ``brane" and ``bulk" channels, as a function of the
cosmological constant and the dimensionality of spacetime. 

Before however writing down the formulae for the energy emission rates, we should first
address the question of the consistent definition of the temperature of an SdS black hole.
Equation (\ref{temp-BH}), for the temperature of the black hole, was derived by making
use of the definition of the surface gravity at the location of the black hole horizon,
i.e.
\beq
k^2_H=-\frac{1}{2}\,\lim_{r \rightarrow r_H}\,(D_M K_N)\,(D^M K^N)\,,
\eeq
where 
\beq
K=\gamma_t\, \frac{\partial \,}{\partial t}
\eeq
is the timelike Killing vector, and $\gamma_t$ a normalization constant. For a static
and spherically-symmetric metric tensor, the above formula reduces to
\beq
k_H=\frac{1}{2}\,\frac{1}{\sqrt{-g_{tt}\,g_{rr}}}\,\Bigl|g_{tt,r}\Bigr|_{r=r_H}\,.
\eeq
In fact, the above expression is derived provided the Killing vector is normalized
according to the rule\,: $K_M K^M=-1$ [for the (-,+,...+) metric signature that  we have
used here] at
the location where $|g_{tt}|=1$. In the case of an asymptotically-flat Schwarzschild
black hole, this happens at infinity where the spacetime reduces to a Minkowski flat
one. However, in the presence of a cosmological constant, this asymptotically-flat
regime does not exist. The normalization condition of the Killing vector can then be
applied at the only point in the Schwarzschild-de-Sitter spacetime that shares the
same characteristic with the infinity in the Schwarzschild spacetime, that is, no
acceleration is necessary for an observer to stay there \cite{Bousso2}. This point,
for a higher-dimensional spacetime, is given in Eq. (\ref{r0}) and corresponds to
the maximum value of the metric function $h(r)$. At the point $r=r_0$, the effect of
the black hole attraction and the cosmological repulsion exactly cancel out. Once
the proper normalization factor is included in the expression of the surface gravity,
the correct expression for the temperature of the black hole comes out to be:
\begin{equation}
T_H = \frac{k_H}{2 \pi} = \frac{1}{\sqrt{h(r_0)}}\,\frac{1}{4\pi r_H}\,
\Bigl[\,(n+1)- \frac{2\kappa^2_D \Lambda}{(n+2)}\,r_H^2\,\Bigl]\,.
\label{temp-new}
\end{equation}

\begin{figure}[t]
	\begin{center}
	\mbox{\includegraphics[scale=0.44]{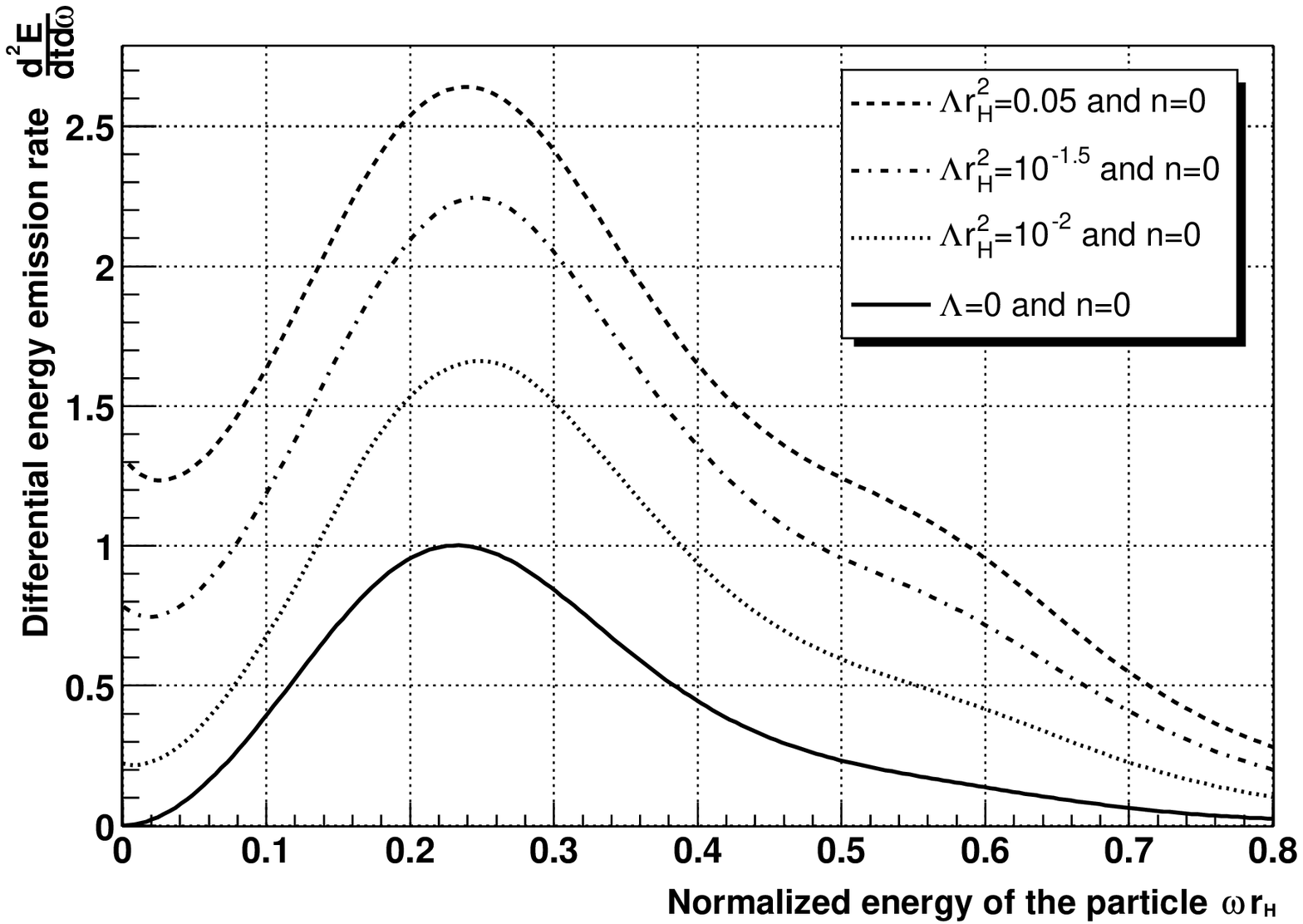}}
	{\includegraphics[scale=0.44]{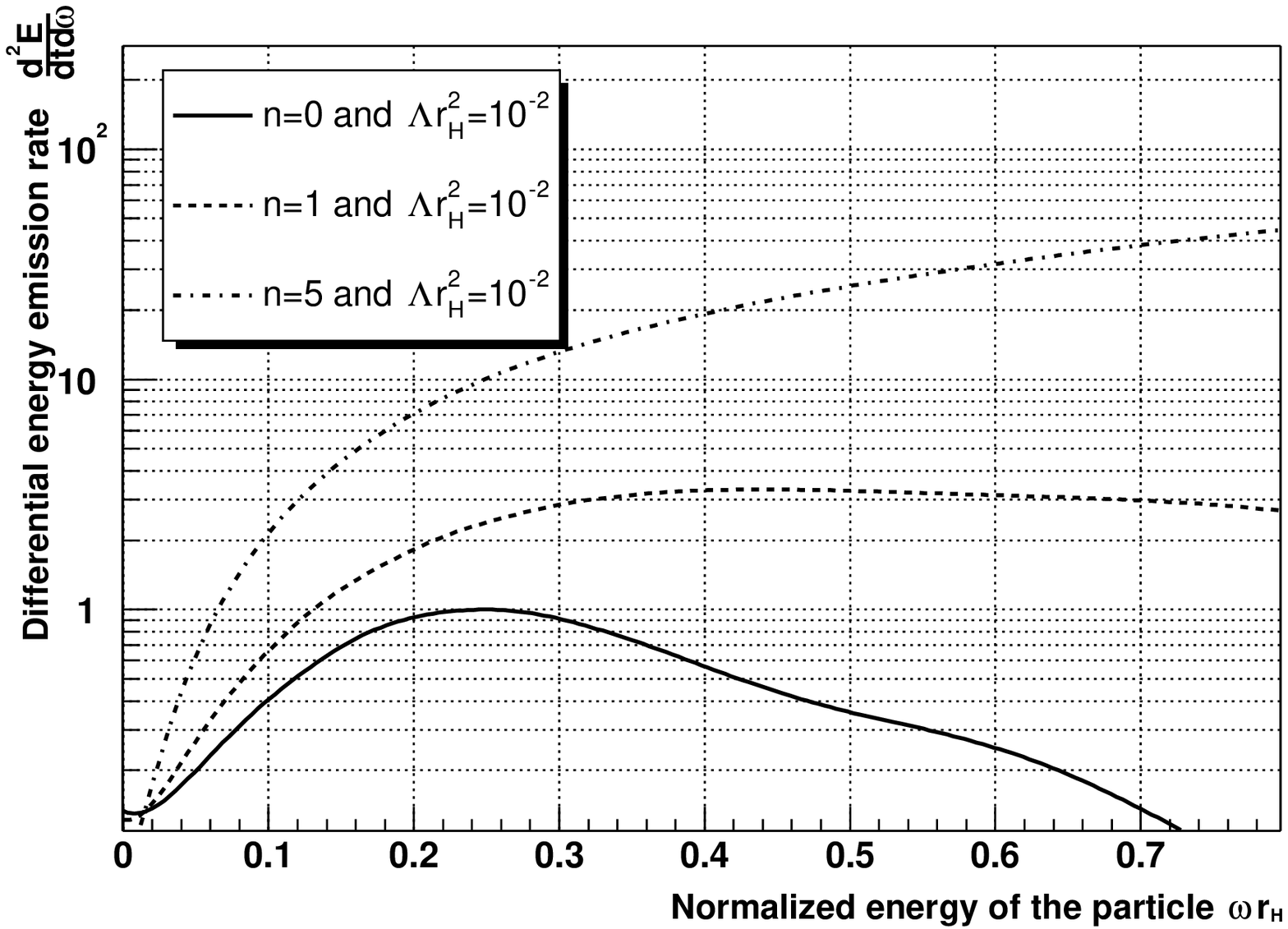}}
	\end{center} \vspace*{-5mm}
	\caption{\it The differential energy emission rate, for scalar emission on
	the brane, versus the dimensionless parameter $\omega{r}_H$: {\rm (a)} the
	dimensionality of spacetime is fixed at $n=0$, while $\Lambda r_H^2$ takes
	the values $\{0,10^{-2},10^{-15},0.05\}$ in Planck units; {\rm (b)}
	the cosmological constant is fixed at $\Lambda r_H^2=10^{-2}$,
	and $n$ takes the indicative values $\{0,1,5\}$.}
\label{flux_br}
\end{figure}

The above formula should then be used for the derivation of the decay radiation spectrum
of a higher-dimensional SdS black hole, both in the bulk and on the brane. The
corresponding expression, Eq. (\ref{power}), takes the form
\beq
\frac{d^2E(\om)}{dt\,d\omega} = \sum_{\ell} N_{\ell,n}\,|{\cal A}_{\ell,n}|^2\,{\om \over
\exp\left(\om/T_{H}\right) - 1}\,\frac{1}{2\pi}\,,
\label{emission}
\eeq
where $N_{\ell,n}=2 \ell+1$, for brane emission, and
\beq
N_{\ell,n}=\frac{(2\ell+n+1)\,(\ell+n)!}{\ell!\,(n+1)!}\,,
\eeq
for bulk emission, and the absorption probability takes the corresponding brane and bulk
values. Figure \ref{flux_br} depicts the differential energy emission
rates for emission on the brane, as a function of both $\Lambda$ and $n$. In order to make
the comparison easier, the maximum point of the lowest curve in both graphs, 5a and 5b,
has been normalized to unity. In Fig. \ref{flux_br}b, where $\Lambda$ is fixed and $n$
varies, the enhancement of the energy
emission rate, by even orders of magnitude as $n$ increases, is clear, and analogous
to the behaviour found in the asymptotically-flat Schwarzschild spacetime \cite{HK}.
A similar enhancement, but of a smaller scale, appears as the value of the cosmological
constant increases. It is worth noting here that in the absence of the normalising
factor $\sqrt{h(r_0)}$, an increase in the value of $\Lambda$ leads to the decrease
of the black hole temperature and thus to a suppression of the energy emission rate;
however, the fact that the normalising factor is always smaller than unity gives a
boost to $T_H$, that overcomes the decrease due to $\Lambda$. 

A distinct feature of the decay spectrum of a SdS black hole is the non-vanishing value
of the emission rate in the limit $\omega \rightarrow 0$. This is caused by the constant
value of the absorption probability in the same limit, which as we saw leads to the
divergence of the greybody factor at the low-energy regime but to a constant value of
the energy emission rate. The larger the cosmological constant is, the larger the value
of the absorption probability, and the larger the number of low-energy states that
are emitted by the black hole. Whereas the total number of states that will be
emitted by the black hole would come as a result of the combined effect of both
$\Lambda$ and $n$, the emission of a significantly number of extremely ultra-soft
quanta on the brane would be a distinctive feature of the spectrum caused by the
presence of a cosmological constant.

\begin{figure}[t]
	\begin{center}
	\mbox{\includegraphics[scale=0.44]{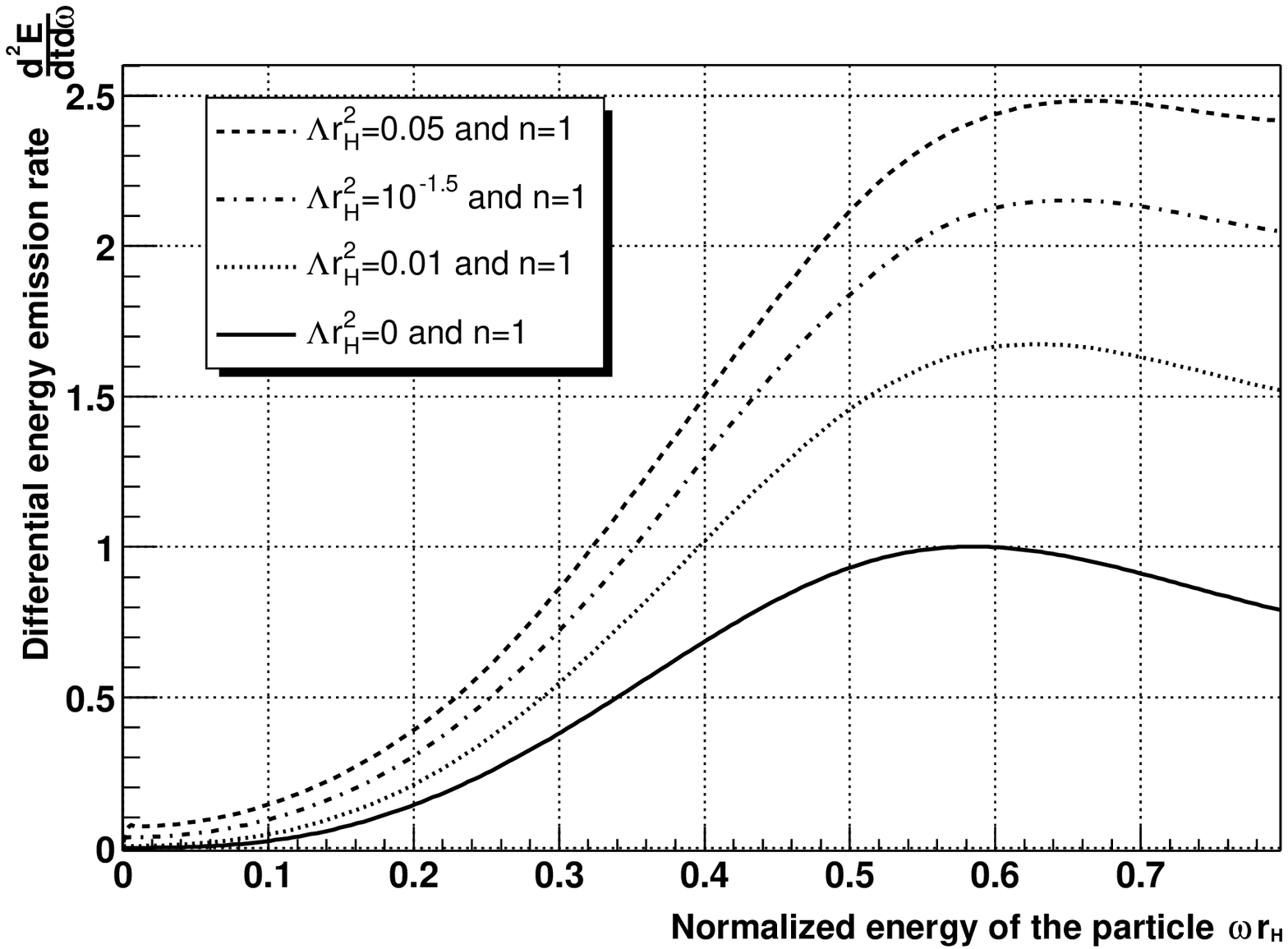}}
	{\includegraphics[scale=0.44]{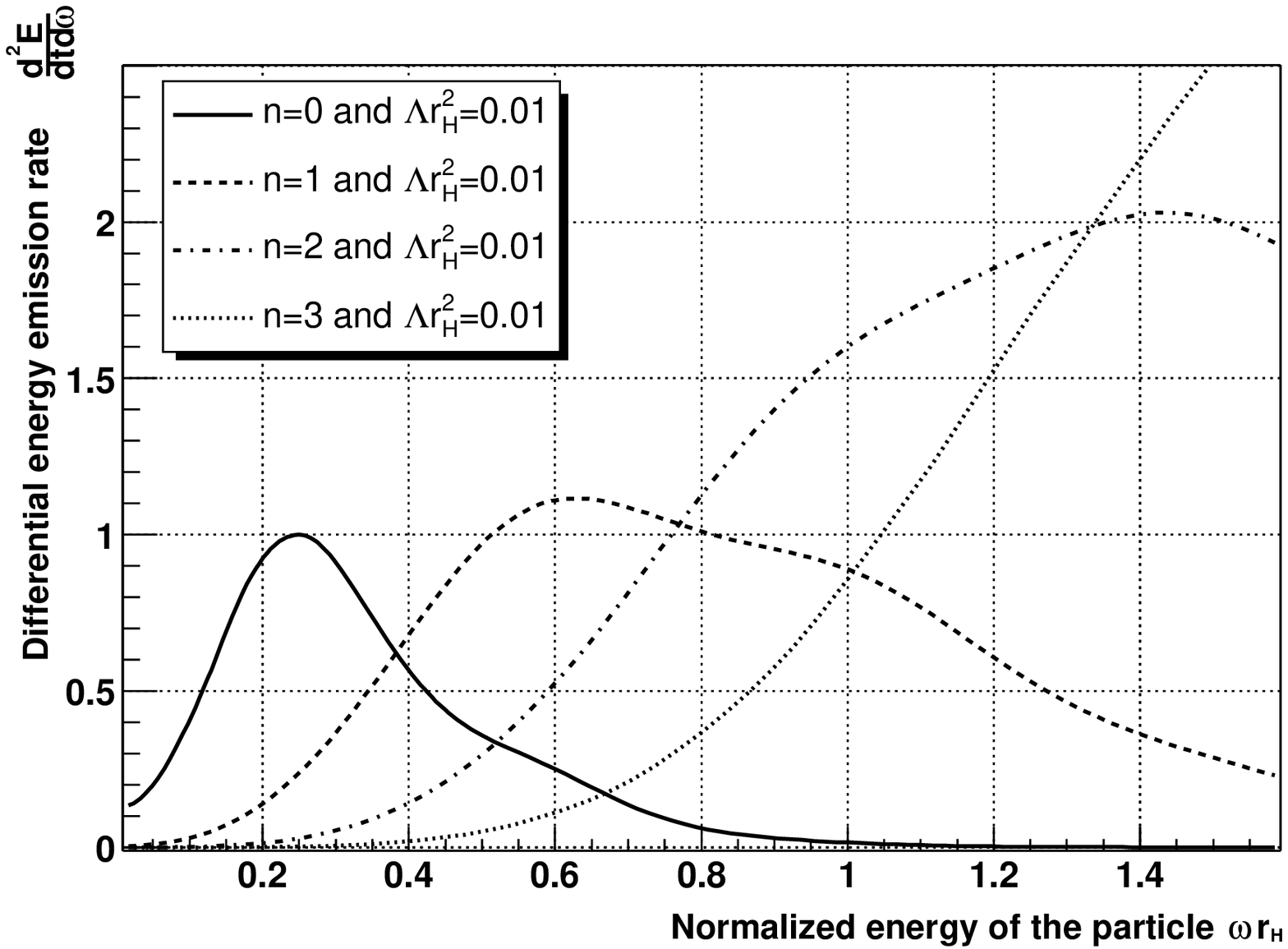}}
	\end{center} \vspace*{-5mm}
	\caption{\it The differential energy emission rate, for scalar emission in
	the bulk, versus the dimensionless parameter $\omega{r}_H$: {\rm (a)} the
	dimensionality of spacetime is fixed at $n=1$, while $\Lambda r_H^2$ takes
	the values $\{0,10^{-2},10^{-1.5},0.05\}$ in Planck units; {\rm (b)}
	the cosmological constant is fixed at $\Lambda r_H^2=10^{-2}$,
	and $n$ takes the indicative values $\{0,1,2,3\}$.}
\label{flux_bulk}
\end{figure}

Figure \ref{flux_bulk} shows the energy emission rates for scalar fields in the bulk
-- the lowest curve in both graphs, 6a and 6b, has been again normalized to unity. As in
the case of brane emission, the bulk energy emission rate is enhanced by the increase
of both the cosmological constant and the dimensionality of spacetime. We notice that,
in both graphs, the peaks of all curves are shifted towards the right, i.e. to higher
values of $\omega r_H$, compared to the ones for brane emission. This is caused by the
much smaller value of the absorption probability at the low-energy regime for emission
in the bulk than on the brane. The total energy available for the emission of particles
by the black hole is given, therefore any suppression in the emission of low-energy
quanta leads to the emission of a larger number of high-energy ones.

\begin{figure}[t]
	\begin{center}
	\mbox{\includegraphics[scale=0.44]{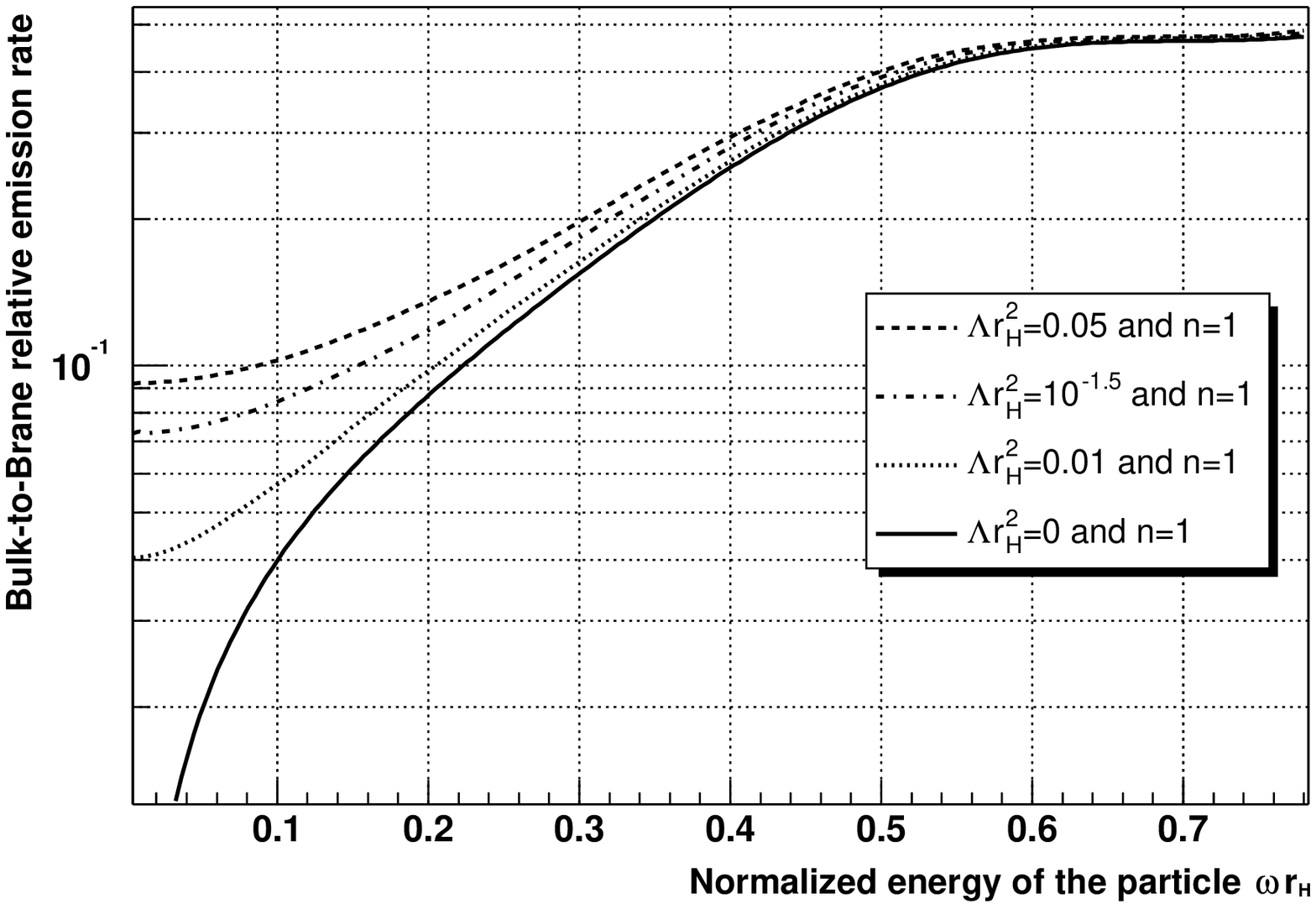}}
	{\includegraphics[scale=0.44]{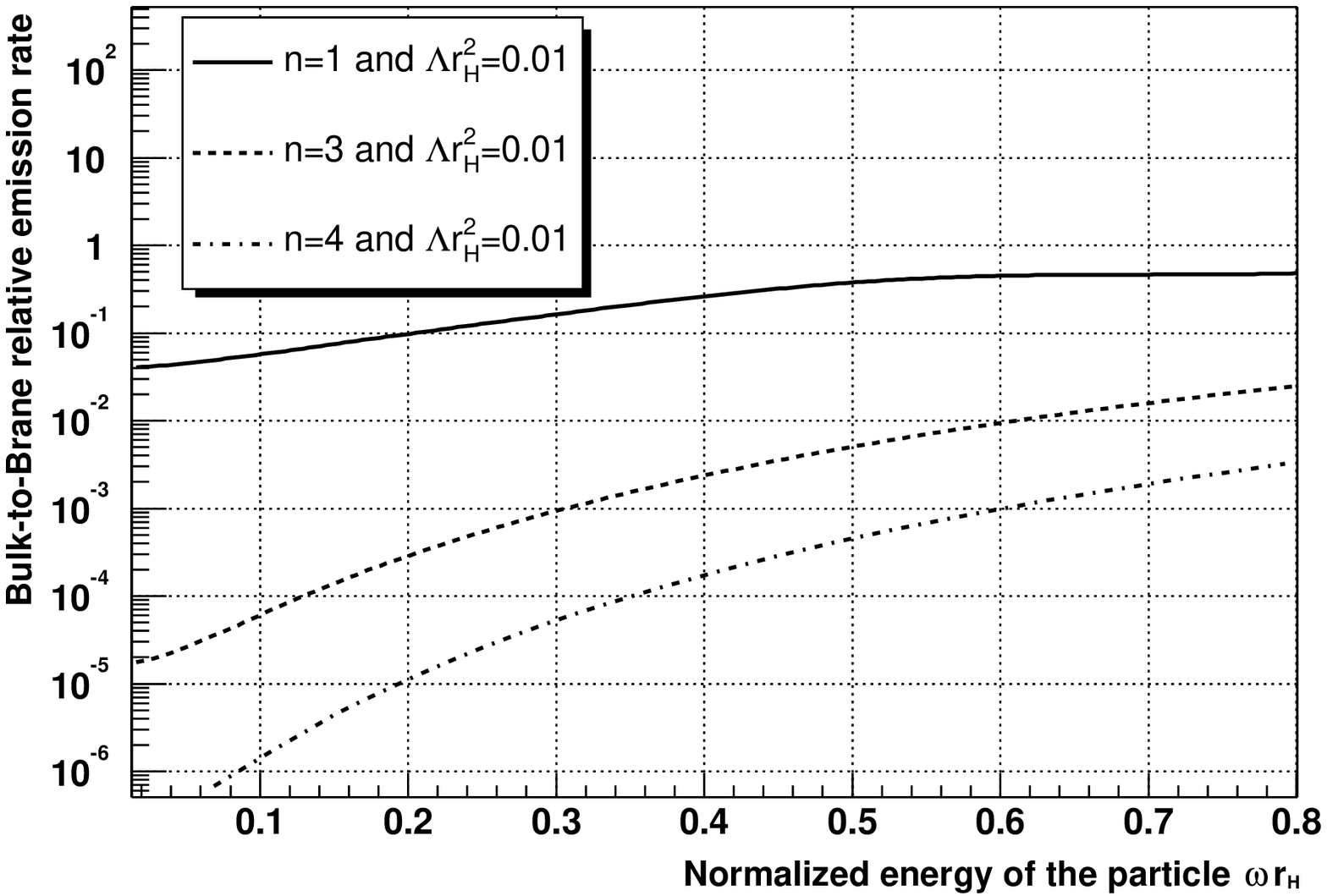}}
	\end{center} \vspace*{-5mm}
	\caption{\it The bulk-to-brane relative emission rate, for scalar
	emission, versus the dimensionless parameter $\omega{r}_H$: {\rm (a)}
	the dimensionality of spacetime is fixed at $n=1$, while $\Lambda r_H^2$
	takes the values $\{0,10^{-2},10^{-1.5},0.05\}$ in Planck units; {\rm (b)}
	the cosmological constant is fixed at $\Lambda r_H^2 =10^{-2}$, and
	$n$ takes the indicative values $\{1,3,4\}$.}
\label{ratio}
\end{figure}

Having derived the exact energy emission rates for emission both in the bulk and on
the brane, we finally turn to the question of the bulk-to-brane relative
emission rate. In the case of radiation emitted by a higher-dimensional Schwarzschild
black hole, it was demonstrated \cite{HK} that most of the energy of the black hole
goes into the ``brane" channel, although the exact emissivity ratio was very much
$n$-dependent. In the case of a SdS higher-dimensional black hole, we need to check
the dependence of that ratio on both $\Lambda$ and $n$; this dependence is shown in
Figs. \ref{ratio}(a,b), respectively. Both figures reveal that this ratio is always
smaller than unity, as in an asymptotically-flat spacetime. The presence of the
cosmological constant -- which can be seen as an addition to the energy of the emitted
particle at every point $r=const.$ according to Eqs. (\ref{eff-eq-br}) and
(\ref{eff-eq-bulk}) -- becomes irrelevant at the high-energy regime, and the curves
match those derived in the Schwarzschild case \cite{HK}. At the low-energy regime,
though, and for fixed $n$ -- see Fig. \ref{ratio}a -- the presence of the cosmological
constant gives a boost to the bulk-to-brane ratio: this is due to the fact that a
non-zero $\Lambda$ lowers the height of the gravitational barrier for a bulk particle
$(n+4)\,(n+2)/8$ times more than for a brane particle. On the other hand, increasing
the number of extra dimensions leads to a strong suppression at the low-energy regime
which, however, becomes milder as the energy increases causing the curves to start
converge at around unity. This effect is obvious in Fig. \ref{ratio}b, and has its
source to the increased number of high-energy modes that are emitted in the bulk,
an effect that becomes more significant for large values of $n$.

\section{Conclusions}

It has by now become clear that the spectrum of Hawking radiation, emitted by a black
hole formed in a general geometrical background, can reveal valuable information for
a variety of parameters that characterize this background. If the black hole is
formed in a higher-dimensional spacetime, the shape and height of the characteristic 
energy emission curves can point towards a particular value of the number of spacelike
dimensions that exist in nature \cite{kmr1,kmr2,HK}. In a similar way, the spectrum
of Hawking radiation emanating from a black hole formed in the presence of the
stringy-inspired Gauss-Bonnet term, is modified when the coupling parameter of this
term varies \cite{Barrau}. Taking this idea further, in this work we have investigated how
the radiation spectrum of a black hole formed in a $D$-dimensional de Sitter spacetime
is affected by the presence of the positive cosmological constant.

In this paper, we have focused on the emission of Hawking radiation from a $D$-dimensional
Schwarschild-de-Sitter (SdS) black hole emitted in the form of scalar fields - the
emission of fields with non-zero spin is studied in a follow-up paper \cite{GKB}. A
higher-dimensional black hole can emit scalar fields both in the bulk and on the brane,
therefore the study of the energy emission rates along both ``channels" must be made.
Scalar particles living on the brane are restricted to propagate on a slice of the
higher-dimensional SdS spacetime, therefore, we had to derive first the induced
gravitational background and then the corresponding equation of motion on the brane.
An analytical calculation led to the derivation of approximate formulae for the
absorption coefficient and greybody factor in the low- and high-energy regime, respectively.
These results served also as guides for the numerical integration, that followed
next and provided us with exact results for the greybody factor valid at all energy
regimes. According to these results, the greybody factor for emission on the brane
has a strong dependence on both the number of spacelike dimensions in nature and the
curvature of spacetime: similarly to the case of an asymptotically-flat black hole,
an increase in the value of $n$ suppresses $\sigma_n(\omega)$; on the other hand, increasing
the value of the cosmological constant enhances the same quantity, especially in the
low-energy regime. By using similar analytical and numerical methods to study the
emission of scalar fields in the bulk, we have found an equally strong dependence of
the ``bulk" greybody factor on $n$ and $\Lambda$, although partially different: here,
both the cosmological constant and a number of additional spacelike dimensions have a
positive effect in the absorption cross-section. 

Computing the final energy emission rates both on the brane and in the bulk was the
following task. We had first to carefully redefine the temperature of the black hole
in such a way as to take into account the non-asymptotic flatness of the SdS
spacetime. This redefinition led to the introduction of an additional normalization
factor whose presence proved to be crucial: in the absence of this factor, the
temperature of the black hole increased with $n$ but decreased with $\Lambda$;
however, in its presence, both parameters caused an increase in the black hole
temperature. Once the exact expressions of the black hole temperature and greybody
factors were used, the energy emission rates in both ``channels" were found to
be strongly enhanced by the presence of either a number of additional dimensions or
a cosmological constant. The rate of enhancement in the energy emission, in terms
of the total number of dimensions and the value of the bulk cosmological constant,
was not the same for the ``brane" and ``bulk" channel. The relative bulk-to-brane
emission rate was then calculated in an attempt to estimate the amount of energy
spent for emission in the bulk -- the space transverse to the brane, and thus
non-accessible to us -- compared to the one spent on our brane. The bulk-to-brane
ratio was found to be both $n$ and $\Lambda$ dependent, as expected, but it remained
smaller than unity for all values of the two parameters considered. As a general
rule, the emission of ``brane" modes dominates in the low-energy regime but at the
high-energy one, the ``brane" and ``bulk" rates become almost comparable.

Apart from the theoretical interest to study effects and processes in a curved spacetime,
motivated by the de Sitter and Anti-de Sitter\,/\,conformal field theory (CFT)
correspondence, there is also a strong observational interest being fuelled by
the prospect of the creation and decay of such small black holes in the near or
far-future collider experiments able to probe the energy regime close to the fundamental
gravitational scale $M_*$. Being able to obtain valuable information on the
topological structure of our spacetime, including its dimensionality and curvature,
from their Hawking radiation spectrum, might be a unique opportunity. According to
our results, not only will the total number of particles and energy emitted by
these small black holes be a function of $n$ and $\Lambda$, but each one of these
parameters will leave a clear footprint in different energy regimes: while the 
cosmological constant becomes almost irrelevant at the high energy regime leaving
the dimensionality of spacetime to determine the spectrum, it gives a unique
feature at the low-energy regime, through a $\Lambda$-dependent constant rate 
of emission of ultra-soft quanta. As scalar
particles have proved to be rather elusive to detect, our present analysis will be
extended, in a follow-up paper \cite{GKB}, to cover the emission of fermions and gauge
fields, too -- the data files used to construct the greybody factors and energy
emission rates, for all species of fields and various values of $n$ and $\Lambda$,
can be found in a publicly accessible web-page \cite{Web}. 

As concluding remarks, let us comment on the validity and usefulness of the results derived
here. The analysis performed in this work is general, and can be applied to any $D$-dimensional
theory with either a low or high gravitational scale $M_*$ -- as long as the
black hole mass remains much larger than this fundamental scale to ensure minimal
backreaction and absence of quantum effects. The number and size of additional dimensions
also remained arbitrary, with the only assumption being made that the black hole horizon
radius is much smaller than the smallest of the compact dimensions. An additional constraint
relevant to the $(4+n)$-dimensional character of our analysis should be now re-instated:
the frequency of the emitted particles must also be larger than $1/L$, otherwise they
should be studied instead in the context of a purely 4-dimensional theory. This is in
complete harmony with the results derived in this work: the effect of the number of extra
dimensions is most obvious in the high-energy regime, where the emitted modes are clearly
(4+n)-dimensional; in the low-energy regime, it is the effect of the cosmological constant
that is the most important, an effect that remains a significant one even in the purely
4-dimensional case, as our results have shown. In the latter case, where no extra dimensions
exist, and bulk and brane coincide, the production of small, classical black
holes in colliders will be impossible, however, the energy emission rates derived
here, and their dependence on the value of the cosmological constant, will still be 
of great use upon the detection of Hawking radiation emitted from small primordial black holes.

\vspace*{0.8cm}
\noindent
{\bf Acknowledgments.} We are grateful to Chris M. Harris for valuable discussions
on the numerical analysis. The work of P.K. was funded by the U.K. Particle Physics
and Astronomy Research Council (PPA/A/S/2002/00350).

\end{document}